\documentclass[manuscript,screen]{acmart}

\usepackage{fixme}
\fxsetup{status=draft}

\usepackage{fontawesome}
\usepackage{xspace}
\usepackage{multirow}

\AtBeginDocument{%
  }

\renewcommand\footnotetextcopyrightpermission[1]{}

\begin{document}

\title{hls4ml: A Flexible, Open-Source Platform for Deep Learning Acceleration on Reconfigurable Hardware}

\author{Jan-Frederik Schulte}
\email{jschulte@cern.ch}
\authornote{Equal contribution.}
\orcid{0000-0003-4421-680X}
\affiliation{%
  \institution{Purdue University}
  \country{USA}
}

\author{Benjamin Ramhorst}
\email{bramhorst@ethz.ch}
\authornotemark[1]
\orcid{0000-0002-0026-1281}
\affiliation{%
  \institution{ETH Zurich}
  \country{Switzerland}
}

\author{Chang Sun}
\email{chsun@cern.ch}
\authornotemark[1]
\orcid{0000-0003-2774-175X}
\affiliation{%
  \institution{California Institute of Technology}
  \country{USA}
}

\author{Jovan Mitrevski}
\email{jmitrevs@fnal.gov}
\orcid{0000-0001-9098-0513}
\affiliation{%
  \institution{Fermi National Accelerator Lab}
  \country{USA}
}

\author{Nicolò Ghielmetti}
\email{nicolo.ghielmetti@cern.ch}
\orcid{0000-0002-4660-9757}

\author{Enrico Lupi}
\email{enrico.lupi@cern.ch}
\orcid{0009-0007-0329-8075}

\author{Dimitrios Danopoulos}
\email{dimitrios.danopoulos@cern.ch}
\orcid{0000-0001-9327-5983}

\author{Vladimir Lon\v{c}ar}
\email{vloncar@cern.ch}
\orcid{0000-0003-3651-0232}
\authornote{Also at Institute of Physics Belgrade.}
\affiliation{%
  \institution{European Organization for Nuclear Research (CERN)}
  \country{Switzerland}
}

\author{Javier Duarte}
\email{jduarte@ucsd.edu}
\orcid{0000-0002-5076-7096}
\affiliation{%
  \institution{University of California San Diego}
  \country{USA}
}

\author{David Burnette}
\email{david.burnette@siemens.com}
\orcid{0009-0003-5033-2599}
\affiliation{%
  \institution{Catapult HLS - Siemens EDA}
  \country{USA}
}

\author{Lauri Laatu}
\email{l.laatu@imperial.ac.uk}
\orcid{0000-0001-6578-8618}
\affiliation{%
  \institution{Imperial College London}
  \country{United Kingdom}
}

\author{Stylianos Tzelepis}
\email{stylianos.tzelepis@cern.ch}
\orcid{0009-0002-8404-6630}
\affiliation{%
  \institution{National Technical University of Athens}
  \country{Greece}
}

\author{Konstantinos Axiotis}
\email{kaxiotis1.0@gmail.com}
\orcid{0000-0003-3664-8186}

\author{Quentin Berthet}
\email{quentin.berthet@hesge.ch}
\orcid{0000-0001-7272-0013}
\authornote{Currently at HEPIA, HES-SO University of Applied Sciences and Arts Western Switzerland.}
\affiliation{%
  \institution{University of Geneva}
  \country{Switzerland}
}

\author{Haoyan Wang}
\email{harry.wang@altera.com}
\orcid{0009-0007-8781-9834}

\author{Paul White}
\email{paul.white@altera.com}
\orcid{0009-0002-7111-2029}

\author{Suleyman Demirsoy}
\email{suleyman.demirsoy@altera.com}
\orcid{0009-0006-5881-7883}

\affiliation{%
  \institution{Altera Corporation}
  \country{USA}
}

\author{Marco Colombo}
\email{mcolom4@uillinois.edu}
\orcid{0009-0008-2446-2898}
\affiliation{%
  \institution{Discovery Partners Institute}
  \country{USA}
}

\author{Thea Klaeboe Aarrestad}
\email{thea.aarrestad@cern.ch}
\orcid{0000-0002-7671-243X}
\affiliation{%
  \institution{ETH Zurich}
  \country{Switzerland}
}

\author{Sioni Summers}
\email{sioni@cern.ch}
\orcid{0000-0003-4244-2061}

\author{Maurizio Pierini}
\email{maurizio.pierini@cern.ch}
\orcid{0000-0003-1939-4268}
\affiliation{%
  \institution{European Organization for Nuclear Research (CERN)}
  \country{Switzerland}
}

\author{Giuseppe Di Guglielmo}
\email{gdg@fnal.gov}
\orcid{0000-0002-5749-1432}

\author{Jennifer Ngadiuba}
\email{ngadiuba@fnal.gov}
\orcid{0000-0002-0055-2935}

\author{Javier Campos}
\email{jcampos@fnal.gov}
\orcid{0009-0008-8029-3267}

\author{Ben Hawks}
\email{bhawks@fnal.gov}
\orcid{0000-0001-5700-0288}

\author{Abhijith Gandrakota}
\email{abhijith@fnal.gov}
\orcid{0000-0003-4860-3233}

\author{Farah Fahim}
\email{farah@fnal.gov}
\orcid{0000-0003-1252-1447}

\author{Nhan Tran}
\email{ntran@fnal.gov}
\orcid{0000-0002-8440-6854}
\affiliation{%
  \institution{Fermi National Accelerator Lab}
  \country{USA}
}

\author{George A. Constantinides}
\email{g.constantinides@imperial.ac.uk}
\orcid{0000-0002-0201-310X}

\author{Zhiqiang Que}
\email{z.que@imperial.ac.uk}
\orcid{0000-0002-9263-6529}

\author{Wayne Luk}
\email{w.luk@imperial.ac.uk}
\orcid{0000-0002-6750-927X}

\author{Alexander Tapper}
\email{a.tapper@imperial.ac.uk}
\orcid{0000-0003-4543-864X}
\affiliation{%
  \institution{Imperial College London}
  \country{United Kingdom}
}

\author{Duc Hoang}
\email{dhoang@mit.edu}
\orcid{0000-0002-8250-870X}

\author{Noah Paladino}
\email{npaladin@mit.edu}
\orcid{0000-0003-1225-537X}

\author{Philip Harris}
\email{pcharris@mit.edu}
\orcid{0000-0001-8189-3741}
\affiliation{%
  \institution{Massachusetts Institute of Technology }
  \country{USA}
}

\author{Bo-Cheng Lai}
\email{bclai@nycu.edu.tw}
\orcid{0000-0002-9729-5196}
\affiliation{%
  \institution{National Yang Ming Chiao Tung University}
  \country{Taiwan}
}

\author{Manuel Valentin}
\email{manuel.valentin@u.northwestern.edu}
\orcid{0000-0002-5298-6225}

\author{Ryan Forelli}
\email{rforelli@u.northwestern.edu}
\orcid{0009-0001-9282-1150}

\author{Seda Ogrenci}
\email{seda@northwestern.edu}
\orcid{0000-0001-8327-9585}
\affiliation{
  \institution{Northwestern University}
  \country{USA}
}

\author{Lino Gerlach}
\email{lino.oscar.gerlach@cern.ch}
\orcid{0000-0002-4153-5541}
\affiliation{%
  \institution{Princeton University}
  \country{USA}
}

\author{Rian Brooks Flynn}
\email{rbflynn@purdue.edu}
\orcid{0009-0006-1662-4506}

\author{Mia Liu}
\email{liu3173@purdue.edu}
\orcid{0000-0001-9012-395X}
\affiliation{%
  \institution{Purdue University}
  \country{USA}
}

\author{Daniel Diaz}
\email{d4diaz@ucsd.edu}
\orcid{0000-0001-6834-1176}

\author{Elham E Khoda}
\email{ekhoda@ucsd.edu}
\orcid{0000-0001-8720-6615}

\author{Melissa Quinnan}
\email{mquinnan@ucsd.edu}
\orcid{0000-0003-2902-5597}

\author{Russell Marroquin Solares}
\email{rmarroquinsolares@ucsd.edu}
\orcid{0000-0002-3364-7463}
\affiliation{%
  \institution{University of California San Diego}
  \country{USA}
}

\author{Santosh Parajuli}
\email{santoshp@illinois.edu}
\orcid{0000-0003-1499-3990}

\author{Mark S. Neubauer}
\email{msn@illinois.edu}
\orcid{0000-0001-8434-9274}
\affiliation{%
  \institution{University of Illinois Urbana-Champaign}
  \country{USA}
}

\author{Christian Herwig}
\email{herwig@umich.edu}
\orcid{0000-0002-4280-6382}
\affiliation{%
  \institution{University of Michigan}
  \country{USA}
}

\author{Ho Fung Tsoi}
\email{hftsoi@sas.upenn.edu}
\orcid{0000-0002-2550-2184}

\author{Dylan Rankin}
\email{dsrankin@sas.upenn.edu}
\orcid{0000-0001-8411-9620}
\affiliation{%
  \institution{University of Pennsylvania}
  \country{USA}
}

\author{Shih-Chieh Hsu}
\email{schsu@uw.edu}
\orcid{0000-0001-6214-8500}

\author{Scott Hauck}
\email{hauck@uw.edu}
\orcid{0000-0001-9516-0311}
\affiliation{%
  \institution{University of Washington}
  \country{USA}
}

\renewcommand{\shortauthors}{Jan-Frederik Schulte, Benjamin Ramhorst, Chang Sun et al.}

\newcommand{\hlsfourml}{\textsc{hls4ml}\xspace}
\newcommand{\kerastwo}{\textsc{Keras~2}\xspace}
\newcommand{\kerasthree}{\textsc{Keras~3}\xspace}
\newcommand{\keras}{\textsc{Keras}\xspace}
\newcommand{\dafourml}{\textsc{da4ml}\xspace}
\newcommand{\torch}{\textsc{PyTorch}\xspace}
\newcommand{\qkeras}{\textsc{QKeras}\xspace}
\newcommand{\brevitas}{\textsc{brevitas}\xspace}
\newcommand{\onnx}{\textsc{ONNX}\xspace}
\newcommand{\qonnx}{\textsc{QONNX}\xspace}
\newcommand{\hgq}{\textsc{HGQ}\xspace}
\newcommand{\np}{\textsc{numpy}\xspace}
\newcommand{\catapulthls}{\textsc{Catapult HLS}\xspace}
\newcommand{\vivadohls}{\textsc{Vivado HLS}\xspace}
\newcommand{\vitishls}{\textsc{Vitis HLS}\xspace}
\newcommand{\oneapi}{\textsc{oneAPI}\xspace}
\newcommand{\quartus}{\textsc{Quartus}\xspace}
\newcommand{\intelhls}{\textsc{Intel HLS}\xspace}
\newcommand{\tensorflow}{\textsc{TensorFlow}\xspace}
\newcommand{\esp}{\textsc{ESP}\xspace}

\begin{abstract}
  We present \hlsfourml, a free and open-source platform that translates machine learning (ML) models from modern deep learning frameworks into high-level synthesis (HLS) code that can be integrated into full designs for field-programmable gate arrays (FPGAs) or application-specific integrated circuits (ASICs).
  With its flexible and modular design, \hlsfourml supports a large number of deep learning frameworks and can target HLS compilers from several vendors, including Vitis HLS, Intel oneAPI and Catapult HLS.
  Together with a wider eco-system for software-hardware co-design, \hlsfourml has enabled the acceleration of ML inference in a wide range of commercial and scientific applications where low latency, resource usage, and power consumption are critical.
  In this paper, we describe the structure and functionality of the \hlsfourml platform.
  The overarching design considerations for the generated HLS code are discussed, together with selected performance results.
\end{abstract}



\maketitle

\section{Introduction}
Recent advances in machine learning (ML) and artificial intelligence (AI) have driven the widespread adoption and deployment of neural networks. For example, cloud vendors are increasingly deploying large language models (LLMs) and deep learning recommendation models as part of their services~\cite{google,aws}.
Similarly, there is an ever-increasing demand for specialized, highly efficient deep learning models for latency- and power-constrained environments, such as real-time systems or edge devices~\cite{Deiana:2021niw}.
Some examples include high-energy physics (HEP) systems at CERN~\cite{CMS-DP-2024-059,CMS-DP-2024-121}, robotics~\cite{aws_robots_1, aws_robots_2}, network infrastructure~\cite{cloudflare}, industrial manufacturing facilities~\cite{aws_siemens}, and satellites and other spacecraft~\cite{edgespaice, nasa}. To tackle the computational and memory requirements of modern neural networks, focus has shifted from conventional CPUs to specialized hardware, such as GPUs, TPUs, and FPGAs~\cite{Decline-CPU}. Complementary to specialized hardware, model compression techniques~\cite{gholami2022survey, cheng2024survey} and novel architectures~\cite{zhou2018graph, hwang2023tutel} are also explored.

In low-latency, real-time systems, FPGAs and ASICs have become the de facto platforms for accelerating neural network, achieving low latency by implementing deeply pipelined designs, specifically tailored to the target model~\cite{nechi2023fpga}. Heterogeneously quantized models~\cite{qkeras, hgq}, or extremely low quantization schemes (e.g.,  binary and ternary models~\cite{Wang_binary, DiGuglielmo_binary}), take advantage of their high configurability to map operations directly to custom logic elements.
FPGAs, in particular, have been deployed in various real-time systems~\cite{CMS-DP-2024-059,CMS-DP-2024-121, microsoft_brainwave, amd_smartcamera, abb_robots}, due to their high flexibility and lower entry barrier compared to ASICs.
However, traditional FPGA development workflows, even when using higher levels of abstractions, such as High-Level Synthesis (HLS), are poorly suited to adapt the functionality of modern deep learning frameworks (e.g., \torch~\cite{torch}, \tensorflow~\cite{tensorflow2015-whitepaper}) as they mainly target CPUs and GPUs. Therefore, deploying neural networks on FPGAs still requires considerable experience and significant time investment. 

Aiming to solve these issues, there has been a growing number of platforms automating neural network deployment on FPGAs (see Section~\ref{sec:related_work}).
While these platforms raise the level of abstraction and represent a significant step forward, each platform provides a range of somewhat arbitrary and application-specific features.
For example, many of these platforms only support models defined in a single deep learning framework, or target only a single type of FPGA~\cite{amd2024vitisai,intel2024fpgasuite,edgecortix2025mera}.
Additionally, the resulting hardware implementations are typically tailored to specific model architectures and precisions, such as convolutional neural networks (CNNs)~\cite{8280160,venieris2016fpgaconvnet,venieris2017latency,venieris2018mapping,9974441,H2PIPE} or multilayer perceptron (MLPs)~\cite{takamaeda2017nngen,dnnweaver:micro16}.
Moreover, these platforms rarely allow the user to choose between different implementations of the models on hardware, which significantly reduces flexibility when balancing latency and resource consumption.
Finally, many of these frameworks~\cite{7966671,takamaeda2017nngen,dnnweaver:micro16} are no longer actively maintained.

In this work, we present \hlsfourml~\cite{fastml_hls4ml}\footnote{Source code and documentation available at \url{https://github.com/fastmachinelearning/hls4ml/} and \url{https://fastmachinelearning.org/hls4ml/index.html}.}, a free and open-source, easy-to-use, and modular platform that translates models from common deep learning frameworks (e.g. \torch, \keras), into low-latency, dataflow designs for FPGAs and ASICs. \hlsfourml acts as a compiler, providing user-facing \emph{front ends} that parse trained models and translate them into an \emph{internal representation} (IR), which is iteratively optimized before hardware lowering through a set of \emph{optimizers}. \emph{Back ends} targeting different HLS compilers map the optimized IR to HLS implementations of layers and operators, and create IP cores for the target model that can be synthesized and integrated into larger applications.

\hlsfourml supports all major deep learning frameworks, including \torch~\cite{torch}, \keras~\cite{chollet2015keras}, and \onnx~\cite{onnx}, as well as their quantized counterparts: \qkeras~\cite{qkeras}, \hgq~\cite{hgq}, \brevitas~\cite{brevitas}, and \qonnx~\cite{Pappalardo:2022nxk, yaman_umuroglu_2023_7622236}.
On the back end side, \hlsfourml supports a growing set of HLS compilers from different vendors, including \vitishls~\cite{vitis} from AMD and \oneapi~\cite{oneapi} from Intel.
In addition to designs for FPGAs, \hlsfourml facilitates the deployment of neural networks on ASICs~\cite{Dickinson:2023yes,parpillon2024smartpixelsinpixelai} through \catapulthls~\cite{CatapultHLS}.
Model architectures that have been implemented using \hlsfourml include MLPs~\cite{Duarte:2018ite}, CNNs~\cite{fastcnn, linebuffer}, recurrent neural networks (RNNs)~\cite{Khoda:2022dwz}, graph neural networks (GNNs)~\cite{Iiyama:2020wap,ds-fpga,Elabd:2021lgo}, and transformers~\cite{Jiang:2024tkg,Jiang:2024lvg}. Additionally, the \texttt{Extension API} allows users to easily add support for any  missing layers or operators, while still leveraging the rest of \hlsfourml's implementations and optimizations.
Configuring the implementation strategy, hardware parallelization, and variable precision, users can optimize their designs for latency or resource usage without modifying or understanding the HLS code directly.
This flexibility allows for rapid prototyping and the co-design of models and hardware.
Finally, \hlsfourml provides multiple environments for model verification.
In addition to conventional software emulation, \hlsfourml enables users to directly deploy their model on specific FPGAs (e.g., Zynq or Alveo cards) with a few lines of Python code.
Owing to \hlsfourml's modularity and standardized interfaces, it is straightforward to integrate it with any platform for FPGA deployment, with out-of-the-box support for commercial~\cite{vitis_kernel, oneapi} or academic~\cite{coyote2} shells.

As a platform, \hlsfourml can be used in two ways: first,
as a deployment platform for various applications, including HEP experiments~\cite{CMS-DP-2024-059,CMS-DP-2024-121}, quantum computing~\cite{Bhat:2024lnc,DiGuglielmo:2025zod}, network firewalls~\cite{ramhorst_dpi}, self-driving cars~\cite{linebuffer}, heart signal monitoring~\cite{10399904}, and space exploration~\cite{edgespaice}.
Second, \hlsfourml can be used as a starting point for research in efficient deep
learning and hardware-model co-design~\cite{que2023metaml, que2025trustworthy, Weitz:2025tcj, que2025metaml, hgq, dspprune, Tsoi:2023isc,symbolnet}. Originally presented in~\cite{Duarte:2018ite} for HEP applications at CERN, \hlsfourml has since grown into a widely used, open-source project for deep learning acceleration and research on custom hardware (see Section~\ref{sec:motivation} for more details on \hlsfourml use cases and applications).
The project is actively maintained with contributions from both academia and industry, is well documented and tested, and frequent supporting tutorials, workshops, and seminars are held to foster the user community.

\section{Background}
\label{sec:background}
FPGAs have emerged as suitable platforms for low-latency, low-power neural network inference, due to their low-level hardware control and high configurability. While development for ASICs is also supported in \hlsfourml, FPGAs are the primary target device and the focus of this paper. In the following, we present a brief overview of common design techniques for model inference on FPGAs. More in-depth overviews are given elsewhere~\cite{nechi2023fpga, boutros2024field,hu_fpga_survey}.

To achieve high throughput and low latency inference, there are many established design techniques. \textit{Parallelization} splits the calculations between multiple processing units (PUs), for example by partitioning the input data between different instances of the layer implementations (data parallelization) or distributing the computation for the neurons of a layer between PUs (model parallelization).
\textit{Pipelining} partitions the calculation in depth so that, for example, subsequent layers are assigned to different PUs, allowing continuous data flow through the design.
This allows the design to accept new inputs before the overall inference is completed, resulting in a low initiation interval~(II).
Most frameworks~\cite{blott2018finn,Duarte:2018ite,9974441, venieris2016fpgaconvnet} use both techniques to achieve maximum performance.

Many frameworks~\cite{finn, H2PIPE, Duarte:2018ite} store weights and intermediate results in on-chip  memory to avoid the overhead of accessing off-chip memory.
These implementations can achieve very low latency and high throughput for models that fit within the limited on-chip logic resources. For larger models, such as transformers, previous studies~\cite{fpga_hbm_nn1, fpga_hbm_nn2, fpga_hbm_nn3} have proposed using high-bandwidth memory (HBM) on recent FPGAs.

For computations, many FPGAs include hardened digital signal processors (DSPs) optimized for multiply-accumulate (MAC) operations with wide bit-widths in deep learning models. Compared to the higher abstraction in CPU and GPU programming, arithmetic operations with these blocks reduce instruction overhead and enable more granular control over data flow. Look-up tables (LUTs) and, sometimes, fast carry chains allow implementing MAC operations without the need for dedicated multiplier blocks like DSPs, either by shift-and-add operations, such as the Booth multiplication algorithm~\cite{booth_multiplication}, or by table look-ups for small bit-width multiplications. Recent works~\cite{PolyLUT, umuroglu2020logicnets} propose training neural networks that can be mapped directly to FPGA LUTs, often achieving high operating frequencies and low resource usage with minimal accuracy loss.

Compared to the commonly used floating-point precisions on CPUs and GPUs, FPGA designs typically represent variables in fixed-precision with lower bit-width. Quantizing a model trained with full precision to lower bit-widths with no or little retraining is known as post-training quantization (PTQ).
Higher degrees of quantization can be achieved while maintaining accuracy by training the model directly with the intended bit-width, a technique known as quantization-aware training (QAT).
Quantization is a method particularly suited for FPGA acceleration, as the arbitrary-precision operation can be efficiently mapped to low-level logic elements in hardware.
Examples of quantization for FPGAs include heterogeneously quantized models~\cite{qkeras, hgq}, as well as binary and ternary models~\cite{Wang_binary, DiGuglielmo_binary, umuroglu2020logicnets}. Additionally, models can also be compressed by pruning parameters, setting some weights or activations to zero.
The level of pruning is selected to balance model size with an acceptable inference accuracy loss~\cite{cheng2024survey, que2025metaml}. On FPGAs, pruning can be structured to be compatible with the low-level hardware implementation~\cite{dspprune}. 

Traditionally, register-transfer level (RTL) design in languages such as VHDL, (System)Verilog, or Chisel have been used to program FPGAs. 
While allowing for more granular control, producing designs in this way is usually more challenging and comes with a steep learning curve. \emph{High-level synthesis} (HLS) abstracts FPGA programming significantly by compiling high-level C/C++ or SystemC code into RTL code for specific target hardware using an \textit{HLS compiler}, such as \vitishls or \catapulthls.
Preprocessor directives, or so-called \emph{pragmas}, are used to guide the compiler in the hardware implementation of the design.
This significantly reduces the difficulty of implementing complex designs for ML inference. While this approach sacrifices some direct control over exact implementations, HLS has been shown to achieve performance on par with hand-written RTL designs~\cite{CarolineICCAD,CarolineThesis} for neural networks (Section~\ref{sec:rtl_vs_hls}).

\section{Overview}
\label{sec:overview}
To achieve high performance, high portability, and easy extensions in the fast-moving field of ML, \hlsfourml is structured in a modular fashion, mimicking modern compilers.
\hlsfourml includes a set of \emph{front ends}, which parse high-level models into an internal representation (IR), a set of \emph{optimizer passes}, which optimize the model graph for the target hardware, and a set of \emph{back ends}, which create high-performance IPs for different flavors of HLS targeting FPGAs from different vendors. Additionally, it includes a rich set of features for model simulation and hardware validation on FPGAs/SoCs.

\begin{figure}
    \centering
    \includegraphics[width=0.7\linewidth]{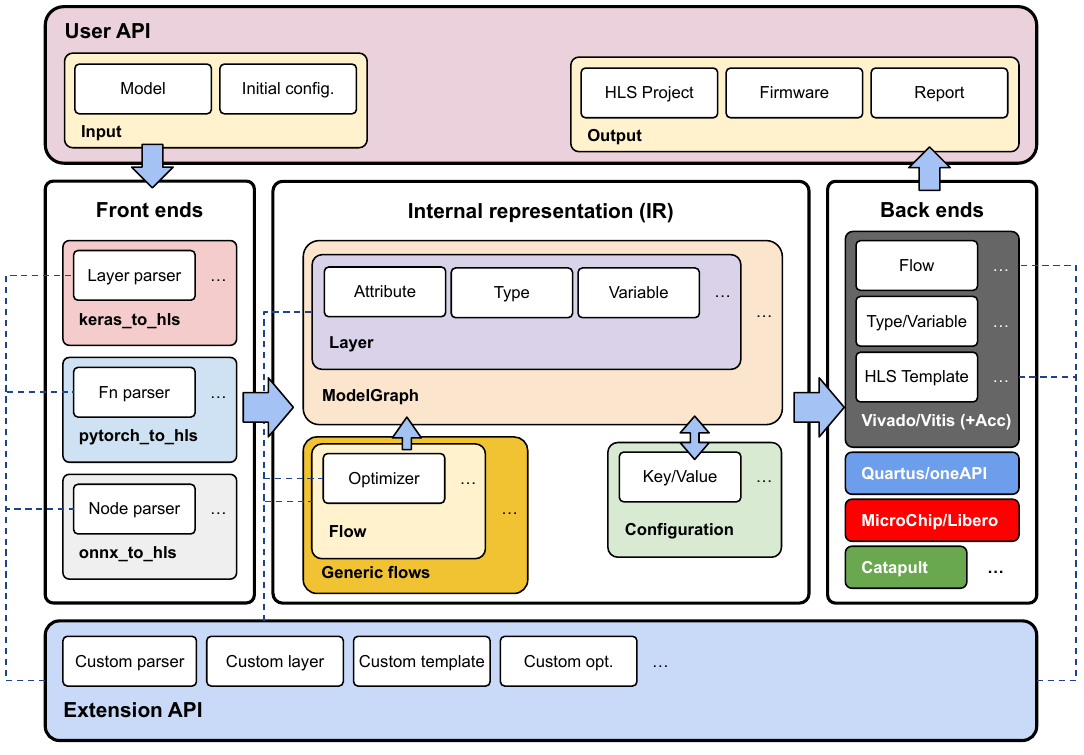}
    \caption{Model conversion and compilation flow in \hlsfourml.}
    \Description{Model conversion and compilation flow in \hlsfourml.}
    \label{fig:strucutre}
    \Description{
        The figure illustrates the structure of the \hlsfourml tool and how it translates an ML model into an FPGA design. On the top, a rectangular box represents the user API. In the middle of the figure there are 3 further rectangular boxes, which from left to right represent the front ends that parse the ML model structure, the \hlsfourml internal representation that translates the model into a standardized format and applies various optimizations to it, and the back ends for the different HLS compilers supported by \hlsfourml.
        Finally, below this there is a further rectangular box for the extension API.
        Arrows between the boxes illustrate the flow of a model, starting in the user API and going through the front ends to the IR, the back ends, and finally back to the user API in form of the output HLS model, firmware, and synthesis report.
    }
\end{figure}

The flow from the initial model to the FPGA IP is illustrated in Figure~\ref{fig:strucutre}.
The users provide a trained model and a configuration file that specifies options such as parallelization factors, variable precisions, and target clock period, which are then parsed by the front-end parsers.\footnote{For more information on the configuration parameters and commands to run these steps, see \url{https://fastmachinelearning.org/hls4ml/api/configuration.html} and \url{https://fastmachinelearning.org/hls4ml/intro/setup.html}.} Generally, \hlsfourml supports fixed-point, exponential, ternary, and binary data types. In the case of PTQ, variable quantization is determined by the precision set in the user configuration, and in the case of QAT, it is automatically inferred from the quantized model (e.g., from \qkeras, \qonnx).
Following that, the model graph is iteratively optimized and refined, given the user configuration and the target back end.
Finally, an FPGA/HLS project containing the generated HLS code and the supporting scripts for simulation, synthesis, and deployment is returned to the user.

\hlsfourml supports all major deep learning frameworks through dedicated front ends for \keras, \torch, and \onnx.
Additionally, \hlsfourml is able to process quantized models from these frameworks; namely models implemented in \qkeras, \hgq, \brevitas, and \qonnx.
We describe the front ends in more detail in Section~\ref{sec:frontends}.
On the back-end side, \hlsfourml keeps the same design principles of modularity and extendability. Currently available are the \textit{Vitis} and \textit{Vivado} back ends for \vitishls, and its deprecated counterpart \vivadohls, for AMD/Xilinx FPGAs, the \textit{oneAPI} and \textit{Quartus} back ends for \oneapi, and its deprecated counterpart \intelhls, for Intel/Altera FPGAs, and the \textit{Catapult} back end for \catapulthls, targeting the Siemens EDA flow.
These back ends are discussed in more detail in Section~\ref{sec:backends}.
To translate between the different front and back ends, \hlsfourml converts the  model into an IR.
The IR initially replicates the structure of the model before a set of optimizers modifies it to better suit the hardware architecture and performance requirements. The IR and optimizers are described in more detail in Section~\ref{sec:IR}. Following a similar modular approach, \hlsfourml implements a library of HLS kernel templates, each implementing the functionality of a model layer or parts thereof. Here, each back end has its own set of templates, which is hand-optimized for the target HLS compiler. To allow greater flexibility and design space exploration different templates are often available that minimize either latency or resources. The user has full control over which of these configurations is used and can also control the degree of parallelism and variable precisions, both of which can be set on a per-layer basis.

Finally, it is important to note that public releases of \hlsfourml support most types of commonly used neural networks.
MLPs and CNNs are fully supported in all front and back ends. RNNs are also supported, though currently not in the \onnx front end.
While some graph neural networks (GNNs) have been realized using \hlsfourml~\cite{Iiyama:2020wap,que2025jedi,ds-fpga}, support for generic GNNs is currently still in development. Specifically, no support for any \torch~\textsc{Geometric} functionality is currently available.
Support for multi-head attention (MHA) transformer models has been introduced in \hlsfourml v1.2.0 via the \hgq~2 front end and is currently only available in the Vitis back end. A detailed overview of the availability of common operators is given in Tables~\ref{tab:support_frontends} and~\ref{tab:support_backends}. In addition, \hlsfourml supports a large number of activation functions, such as ReLU or softmax, as well as general tensor operations (e.g. reshaping, concatenation, or addition).
Even though many layers are supported, a model may not necessarily be suitable for acceleration with \hlsfourml. Very large or complex models can fail to pass HLS compilation due to the on-chip, low-latency dataflow implementation of \hlsfourml designs which can lead to high resource utilization and congestion (more details on the limitation of \hlsfourml can be found in Section~\ref{sec:limitations}). Such models should be compressed through quantization and pruning, or, in some cases, consider alternative acceleration techniques, leveraging for e.g., overlay architectures or off-chip memory.
If a model contains currently unavailable operations/layers, \hlsfourml offers an \texttt{Extension API} to add support for additional layers (Section~\ref{sec:extension}), while still relying on the implementations for existing layers, optimizers, and scripts.

\begin{table}[htb]
    \newcommand{\yes}{\faCheckSquare}
    \newcommand{\no}{\faSquareO}
    \newcommand{\partially}{\faPlusSquareO}
    \centering
    \caption{Support for different NN layers or types in the \hlsfourml front ends. Support is indicated by \yes, while not supported layers are marked \no. Layers not supported in the respective NN frameworks are marked \---.}
    \begin{tabular}{c|cccccccc}
        \multirow{2}{*}{Layer} & \multicolumn{8}{c}{Front end}                                                                   \\
                               & \kerastwo                     & \qkeras & \hgq & \kerasthree & \hgq~2 & \torch & \onnx & \qonnx \\ \hline
        Linear/Dense           & \yes                          & \yes    & \yes & \yes        & \yes   & \yes   & \yes  & \yes   \\
        1D/2D Convolution      & \yes                          & \yes    & \yes & \yes        & \yes   & \yes   & \yes  & \yes   \\
        LSTM/GRU                   & \yes                          & \yes    & \--- & \yes        & \---   & \yes   & \no   & \no    \\
        Einsum                 & \no                           & \---    & \--- & \yes        & \yes   & \yes   & \no   & \---   \\
        Multihead Attention    & \no                           & \---    & \--- & \no         & \yes   & \no    & \no   & \---   \\
        Batch Normalization    & \yes                          & \yes    & \yes & \yes        & \yes   & \yes   & \yes  & \yes   \\
        Layer Normalization    & \yes                          & \no     & \no  & \no         & \no    & \yes   & \no   & \no    \\
        GNN                 & \no                          & \no    & \no  & \no        & \no   & \no   & \no  & \no   \\
    \end{tabular}
    \label{tab:support_frontends}
\end{table}

\begin{table}[htb]
    \newcommand{\yes}{\faCheckSquare}
    \newcommand{\no}{\faSquareO}
    \newcommand{\partially}{\faPlusSquareO}
    \centering
    \caption{Support for different layers in the \hlsfourml back ends. Support is indicated by \yes, while not supported layers are marked \no.}
    \begin{tabular}{c|ccc}
        \multirow{2}{*}{Layer} & \multicolumn{3}{c}{Back end}                             \\
                               & Vitis/Vivado                 & Quartus/oneAPI & Catapult \\ \hline
        Linear/Dense           & \yes                         & \yes           & \yes     \\
        1D/2D Convolution      & \yes                         & \yes           & \yes     \\
        LSTM/GRU                  & \yes                         & \yes           & \yes     \\
        Einsum                 & \yes                         & \no            & \no      \\
        Multihead Attention    & \yes                         & \no            & \no      \\
        Batch Normalization    & \yes                         & \yes           & \yes     \\
        Layer Normalization    & \yes                         & \no            & \no      \\
        GNN                 & \no                         & \no            & \no     \\
    \end{tabular}
    \label{tab:support_backends}
\end{table}

\section{Front ends for different ML libraries}
\label{sec:frontends}
The front end converts models from the supported frameworks mentioned above into the IR.
For each deep learning framework, there is a dedicated front end that (1) parses the model graph, and (2) extracts relevant information from each individual layer or operation.
The second step is implemented with an extensive repository of layer handlers, one for an individual, or a family of supported layers or operations.
Each layer handler accepts a layer object and returns a dictionary with the necessary configuration and data for representing the layer in \hlsfourml's IR.
All weights in the model are converted to \textsc{NumPy} arrays at this stage, and front-end specific objects are eliminated for compatibility.

\subsection{Keras}
\label{subsec:keras}

Traditionally, \hlsfourml has focused on models created in \kerastwo and supports the most commonly used layers in this framework.
For \kerastwo, both Keras objects and serialized representations of the model, which contain detailed information about the model architecture and layer configurations, are accepted.
This allows the conversion of models that are saved in the \texttt{HDF5} format without requiring the Keras library to be installed. Direct parsing of \texttt{.keras} files is not supported due to the absence of a public API for reading model configurations. However, such models can be loaded in Keras and passed to \hlsfourml as in-memory objects. \kerasthree introduces major changes to the \textsc{Keras} library. Therefore, the \kerasthree parser does not inherit from the \kerastwo parser.
The \kerasthree parser directly ingests a \texttt{keras.api.Model} object, and reconstructs the model architecture via data dependency solving.
\kerasthree layer handlers also ingest \texttt{keras.api.layers.Layer} objects directly, and return a dictionary with the layer configuration and weights.
When a layer type is not supported, the parser will attempt to fall back to the \kerastwo layer handler (if available) to extract the configuration.

The \kerastwo parser supports models quantized by two different QAT libraries, \qkeras and \hgq, while the \kerasthree parser currently only supports \hgq.
For quantized layer types, the quantization parameters are included in the IR of the quantized layer configuration. Quantization parameters derived from QAT libraries are enforced in \hlsfourml and will override any user-provided configurations.
\qkeras supports three major quantization types: fixed-point integer based, binary, and power-of-two quantization, of which are all supported by \hlsfourml, while the power-of-two quantization may only be used for the weights.
For \hgq models, a dedicated optimizer pass in the IR is invoked for precision configuration.
This involves symbolic precision propagation through the model, and all quantization parameters are derived from explicit quantization layers converted from \hgq, along with the weights provided.
User-provided precision configurations are discarded when converting from \hgq, and conversions from properly defined \hgq models are always bit-exact up to the floating point representation limits.

\subsection{\torch}
Symbolic tracing in the \textsc{torch FX} framework is used to infer the architecture and configuration of \torch models.
In general, \hlsfourml supports operations implemented as both \texttt{torch.nn.Module} objects and stateless operators in the \texttt{torch.nn.functional} module.
The support for \torch layers generally matches that of \kerastwo. 
A custom tracer inheriting from the \textsc{torch.fx.Tracer} class is used to enable the parsing of custom layers in the \texttt{Extension API} (Section~\ref{sec:extension}). Since \hlsfourml was first developed to support \kerastwo models, which adopts the ``channels-last'' convention for multi-dimensional tensors, many HLS kernel templates only have a ``channels-last'' version implemented. \torch uses the ``channels-first'' convention in most of its operations.
Therefore, an optimizer adds the necessary transpose operations to the \hlsfourml IR to ensure compatibility with the ``channels-last'' convention.

Models trained using QAT in \brevitas can be converted to the \qonnx format using the \brevitas-\onnx export functionality, allowing them to be ingested by the \onnx front end described below. This approach is also common in other frameworks~\cite{finn, blott2018finn}. In addition, support for direct parsing of \brevitas models, without reliance on \qonnx, as part of the \torch front end is in development.

\subsection{\onnx}
\begin{figure}[htpb]
    \centering
    \includegraphics[width=0.5\linewidth]{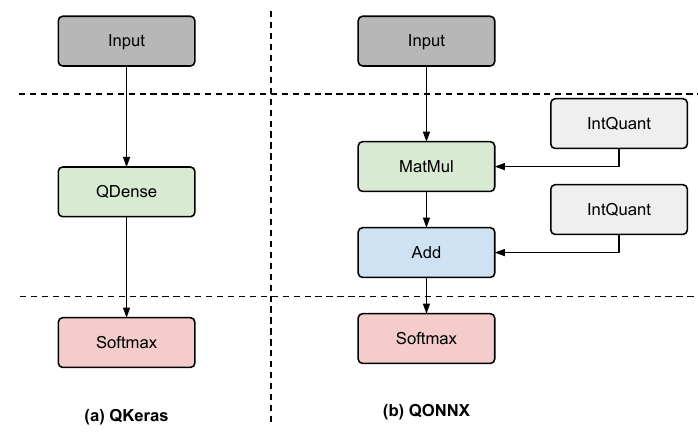}
    \caption{Comparison of the representation of an example model consisting of a one-layer MLP with a softmax output layer in \qkeras (left) and \qonnx (right).}
    \label{fig:qonnx}
    \Description{
        Comparison of the structure of a simple one-layer model consisting of an input, a Dense layer, and a SoftMax activation, in QKeras (left) and QONNX (right). The layers are represented in colored rectangles, connected with a dashed line representing the flow of data between them, and arranged in a vertical line so that the data flows from top to bottom. In the QKeras version on the left, there are three rectangles, for the Input, the QDense layer, and the Softmax. In the QONNX version on the right, the Dense layer is split into a MatMul operation and a Add operation. For both of these, there are additional rectangles to the top-right representing the Quant operation in QONNX, which are connected to the MatMul and Add layer with arrows indicating that the Quant operations are applied to them.
    }
\end{figure}

\onnx models are supported with direct object parsing.
The converter ingests an \onnx object and parses its graph to determine the model inputs, outputs, and all operators present in the model. Because of the fine-grained representation of operators in \onnx models, \hlsfourml requires the model to be preprocessed, usually referred to as ``cleaning'', before conversion.
Cleaning is performed using the \qonnx~\cite{yaman_umuroglu_2023_7622236} package, which performs constant folding, shape inference, renaming of the nodes into human-readable names, and conversion of the graph into the ``channels-last'' convention.
Additionally, \qonnx, as an extension of the \onnx library, introduces three new operators: \textsc{Quant}, \textsc{BipolarQuant}, and \textsc{Trunc} to enable flexible representation of quantized models. \hlsfourml supports a subset of models representable by \qonnx that are physically feasible for FPGA implementation.
In the \hlsfourml IR, the precision is derived from the quantization operators and enforced.
Additional scaling or unscaling is applied as needed.
Figure~\ref{fig:qonnx} illustrates a QONNX representation of a simple one-layer MLP with a softmax output layer, compared to the representation in \qkeras.

\section{Internal representation and optimizer flows}
\label{sec:IR}
\subsection{Internal Representation}
The \hlsfourml IR aims to provide a front- and back-end agnostic representation of models. By abstracting models from different frameworks into a uniform representation, \hlsfourml can apply transformations and optimizations to a model independent of the framework used to create it. The IR is implemented as a \texttt{ModelGraph} object where each node corresponds to a layer or operator in the network. Each node contains all layer-specific information such as operation type, weights, quantization method, and connections to other nodes.
During conversion, the IR is initialized with the layers from the original model by the front-end converter. Then, the IR undergoes a series of transformations performed by the \emph{optimizer flows} to reach the final state used in the code generation of the target back end.

By default, \hlsfourml creates a single HLS project for the entire model, which is then synthesized into a monolithic design. As of v1.2.0, \hlsfourml supports a new feature that allows for the splitting of the model into multiple subgraphs. This is implemented by extending the \texttt{ModelGraph} with the \texttt{MultiModelGraph} object, which splits the model at user-defined layers, and creates independent HLS projects that can be synthesized independently. This allows for parallel synthesis to reduce synthesis time, or potentially distributing large models over multiple chips or devices in the future. \hlsfourml currently supports automatic stitching of the synthesized IP cores from subgraphs into a single design for the Vivado/Vitis back ends. By running the individual synthesis in parallel, the HLS synthesis time for a quantized ResNet for CIFAR-10 classification decreases from 7\,h to 3\,h on average.

\subsection{Optimizer flows}
Inspired by modern compilers, \hlsfourml implements a series of transformations, called \emph{optimizer flows}, that iteratively transform the IR for the target back ends, where each flow represents a distinct optimization stage.
Optimizer flows can either be generally applicable to all models or specific to a target back end. A single flow implements multiple \emph{optimizers}, each performing a specific transformation on the model, such as precision propagation, template instantiation, or removing redundant operations. For instance, an optimizer may fuse batch normalizations that immediately follow an affine projection (fully connected or convolutional layers) operation into a single node with equivalent, fused weights when both nodes are not quantized.

\subsection{User directives}

In addition to the graph representation of the model itself, the \texttt{ModelGraph} object also holds directives regarding the model conversion and optimization process in the form of an \texttt{HLSConfig} object. The \texttt{HLSConfig} object contains information that cannot be derived directly from the model, such as the target back end, implementation strategy, I/O type, and, in some cases, quantization precision and resource reuse preferences. Users provide these configurations as a Python dictionary. Some of the hardware-oriented configurations (e.g., I/O type, strategy) are further discussed in Section~\ref{sec:hardware}, in conjunction with the hardware implementation of neural network kernels.

Quantization parameters are a crucial part of the model conversion process. Improperly configured quantization parameters may introduce extra errors that accumulate throughout the model, which can lead to significant performance degradation. 
Currently, \hlsfourml supports fixed-point, exponential (power-of-two), ternary, and binary data types. When the front-end framework quantizes the model properly, the quantization parameters will be stored directly in the nodes. However, some precision information may not always be available depending on the front end used. For instance, the accumulator\footnote{The accumulator holds the intermediate results in a layer for the MAC operations.} precision is rarely available from the front-end libraries. Since v1.0.0, \hlsfourml is able to automatically determine it through conservative estimation to avoid unwanted overflows when the user specifies \texttt{"auto"} for the accumulator precision in the configuration.
Other precision types, such as the weight, bias, and result precision, can also be inferred automatically, depending on the layer type and the metadata it contains when \texttt{"auto"} is specified.
Since v1.2.0, \hlsfourml includes a dedicated optimizer pass that propagates precision at the model level for bit-exactness when the model is fully quantized, which relies only on explicitly defined quantizers and the weights provided through interval arithmetic.
User-supplied precision configurations are ignored when this optimizer pass is enabled.

\section{Back ends and hardware implementations}
\label{sec:backends}

\hlsfourml generates HLS code through a set of manually optimized templates for different layers and operations, which are combined to represent a complete model to be synthesized.
Each back end includes a set of dedicated templates and optimizers. Some back ends, such as those for \vivadohls and \vitishls, may share some templates and optimizers because of their similarity.
The optimizers will register the back-end specific templates and apply them to the IR. They also perform back-end specific optimizations, such as pragma insertion, precision propagation, constant folding, and others.
When creating an HLS project, the back ends will also generate the necessary utility files such as TCL scripts for synthesizing the project, test benches, and bridge code for calling the generated libraries directly from high-level Python functions.
After synthesizing the generated HLS code, the back ends can parse the synthesis reports and display a concise summary of the resource usage and latency.

\subsection{Hardware implementations}
\label{sec:hardware}
In the following, we provide a brief overview of hardware implementations in \hlsfourml, with an emphasis on constant matrix-vector multipliers (CMVM) and activations, with are core building blocks of neural networks. Further details on specific layer implementations can be found in the corresponding publications for MLPs~\cite{Duarte:2018ite}, CNNs~\cite{fastcnn, linebuffer} and RNNs~\cite{Khoda:2022dwz,quartus_rnn_Aad2021}. In addition, we discuss the various strategies and data flow options supported in \hlsfourml, each of which provides a trade-off between the resources used and inference latency.

\begin{figure}
    \centering
    \includegraphics[width=\linewidth]{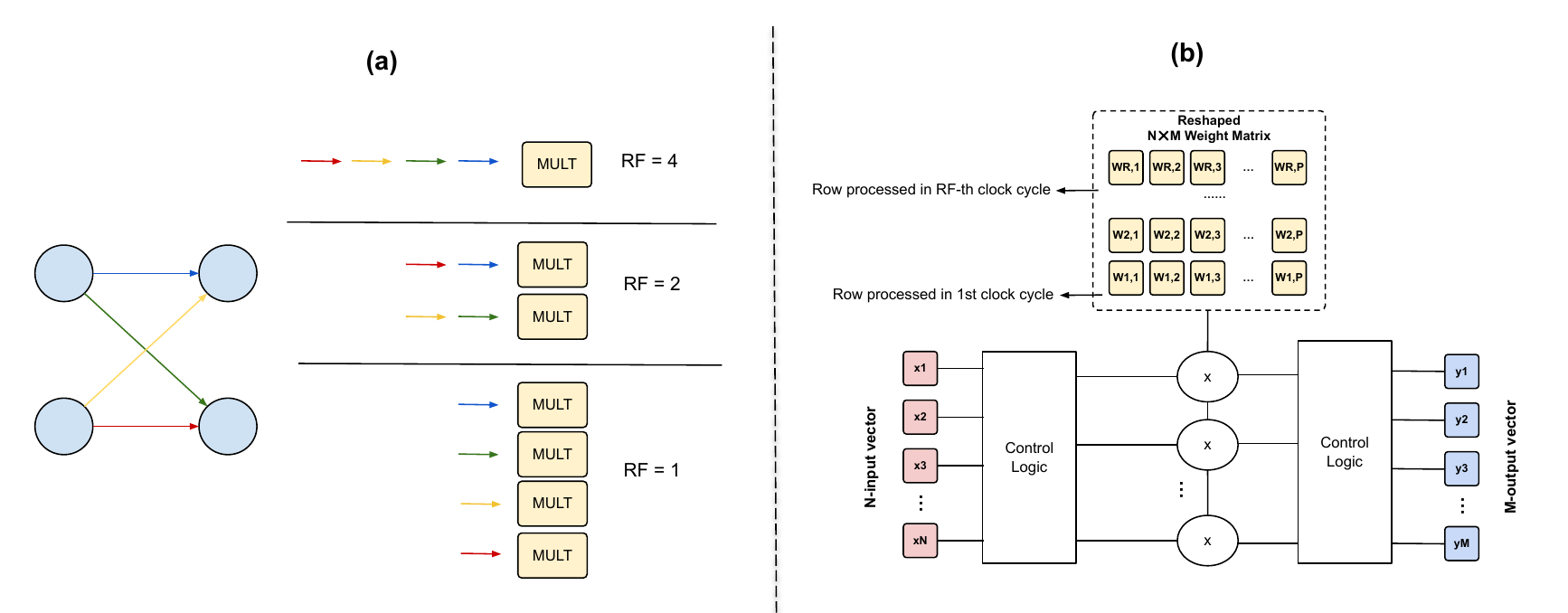}
    \caption{
        (a) Illustration of the effect of different RF values for the outer product of two two-vectors. (b) Example CMVM in \hlsfourml with the Resource strategy.
        Given a linear layer with $N$ inputs, $M$ outputs and reuse factor RF, there will be $P = \frac{M \cdot N}{\text{RF}}$ multipliers operating in parallel. In each clock cycle, the control logic selects $P$ out of the $N$ inputs and feeds them to the multipliers, with wrap around if $P > N$.
        The $N \times M$ kernel is reshaped and mapped to on-chip memories such that $P$ elements can be accessed in parallel in each clock cycle.
        The products are accumulated accordingly at the precision specified to form the output.
    }

    \label{fig:resource_cmvm_example}
    \Description{
        (a) The left figure illustrates the trade-off between the resources (multipliers) used and the latency for varying reuse factors (RFs). For a 2 x 2 input vector, the top case indicates the case when RF equals 4 and one multiplier is used across four clock cycles, leading to high latency but lower resource usage. The middle case indicate the case when reuse factor equals to 2 and the multiplication is implemented using 2 multipliers across two clock cycles. Finally, the bottom cases corresponds to the case when the reuse factor is equal to 1; corresponding to all multiplications executed in parallel with 4 multipliers, leading to the lowest latency but highest resource usage. (b) Schematic view of a constant matrix-vector multiplication operation as implemented on FPGA using hls4ml.
        In the top middle, a rectangular area represents the weight tensor, with smaller yellow square inside that represent the individual weights.
        The matrix has been reshaped for implementation in registers or BRAM to have a shape of P times RF, where P is the number of multipliers used in the matrix multiplication and RF is the reuse factor chosen for this implementation.
        The lower half of the figure illustrate the multiplication, with a vertical row of red squares on the left representing the input vector, which is connected to a rectangle for the control logic to the right of it, followed by a vertical row of circles that represent the multiplication, another rectangle for the control logic for the accumulation, and finally a vertical row of blue squares for the output vector.}
\end{figure}

\paragraph{Constant matrix-vector multiplication, strategies, and parallelism:} The CMVM operation is a fundamental building block of neural networks. \hlsfourml places significant emphasis on its efficient implementations, with three supported strategies: \emph{Latency}, \emph{Resource}, and \emph{Distributed Arithmetic}. To control the degree of parallelism, \hlsfourml allows users to set the \textit{Reuse Factor} (RF). Given a weight matrix of size $M \times N$ and a reuse factor $RF$, \hlsfourml implementations of CMVM are such that there are no more than $N\_MULT = \frac{M \cdot N}{RF}$ multiplications per clock cycle (e.g., Fig.~\ref{fig:resource_cmvm_example}a).
Different strategies will differently impact latency and resource consumption of a layer, and users are advised to choose based on their design goals. The availability of a particular strategy for a layer varies across back ends.

The \emph{Latency} strategy implements fully unrolled CMVM with weights directly embedded into multiplier circuits, allowing the HLS compiler to perform intensive optimizations, such as removing multiplications with zeros, replacing multiplications with specific constants with shift-and-add operations, or reordering the multiplications to improve timing. The RF is implemented as a compiler directive that limits the number of parallel multiplications and determines the initiation interval of the CMVM operation. The HLS compiler determines the exact ordering and hardware primitive for the MAC operations, based on the data type, precision, and other conditions (e.g. target hardware and frequency).
In cases where DSP blocks or other hardened multipliers are used for the MAC operations, a larger RF will usually result in reusing those blocks and lower resource consumption.
However, when MAC operations are implemented in fabric, such as LUTs and fast carriers with shift-and-addition operations, a higher RF will not necessarily result in resource reduction, but will still increase the II. Depending on the underlying compiler and hardware mappings, the resource usage may even increase. Therefore, the Latency strategy is suitable for small models, targeting very low inference latencies. Additionally, the RF should be tuned carefully as it may not yield resource savings and may increase II.

In the \emph{Resource} strategy, the CMVM implementation stores the weights in BRAM and explicitly controls the degree of parallelism in by specifying the number of MAC operations performed in parallel per clock cycle (Fig.~\ref{fig:resource_cmvm_example}a). Weights stored in BRAM are partitioned into $RF$ blocks, each of size $N\_MULT$ (Fig.~\ref{fig:resource_cmvm_example}b). Each clock cycle, a slice of $N\_MULT$ weights is accessed and processed together with the inputs by a MAC unit, leading to an II of $RF$ clock cycles. Depending on the chunk size $N\_MULT$, the weights may be stored in multiple BRAMs to increase the number of ports for parallel read. The exact implementation of the MAC unit will depend on the bit-width of the weights and inputs, as well as compiler heuristics. In the case of DSP-based MAC units, it can be implemented as cascaded DSP tiles, whereas for LUT-based MAC units, the exact implementation differs heavily depending on the precisions, target frequency and other factors. However, for specific operand types, \hlsfourml can overwrite compiler primitives; for e.g. multiplications with power-of-two weights can be implemented using shifters explicitly, while multiplications with binary and ternary weights can be realized with AND and XOR gates. Increasing the RF in the Resource strategy directly reduces the number of MAC units used, thus reducing resource consumption. The Resource strategy is better suited for large models, since the Latency strategy can fail to compile in these cases due to full loop unrolling. However, due to the overhead in data movement and the use of general MAC units, the end-to-end latency for the Resource strategy is typically higher than for the Latency strategy with the same RF.

The \emph{Distributed Arithmetic} (DA) strategy implements the CMVM operation by explicitly decomposing it into an adder graph with only shift-and-add (or subtract) operations, where any bit-wise sparsity in the weights will be explicitly exploited.
This strategy relies on an external library, \dafourml~\cite{da4ml}, for generating and optimizing the adder graph corresponding to the specific CMVM problems. Compared to the Latency strategy with an RF of one, the DA strategy usually has lower fabric resource utilization and latency. DSPs or other hardened multipliers will not be used due to the lack of explicit multiplications.
However, the DA strategy in general does not support RFs greater than one since the generated adder graph effectively unrolls the entire CMVM operation.
Hence, in case of moderately large CMVM problems consuming significant amount of hardened multipliers (e.g., DSPs), the user is encouraged to try the Latency or Resource strategy with a higher RF instead of the DA strategy.

Some layers, e.g., convolutional layers lowered to CMVM through the im2col~\cite{im2col} transformation, may perform identical CMVM operations multiple times on different inputs in one forward pass. In this case, the parallelism between CMVM operations is controlled by the \emph{Parallelization Factor} (PF).
The PF controls the number of CMVM operations that are performed in parallel, and it is independent of the strategy and RF used for the CMVM operation. A higher PF reduces the II of one layer at the expense of higher resource usage.
Like RF in the Resource strategy, the PF chosen must fully divide the number of CMVM operations needed in one forward pass for the layer to avoid any remainder.
The total II of one layer is determined by the PF (if available), RF, and strategy together.

\paragraph{Activations:} 
    Piecewise linear activations (e.g., ReLU, Leaky ReLU) are implemented using multiplexers. Other activations (e.g., tanh, sigmoid) are implemented as  lookup tables\footnote{Not to be confused with the LUT primitive on FPGAs.}, which are populated at compile-time. The use of look-up tables guarantees constant and efficient calculation of the activation, regardless of its complexity. Given the precision of the layer input, hls4ml will generate a look-up table storing all possible output values of the activation function for each possible input value. In case the user-specified table size is smaller than the generated table, \hlsfourml will drop the least significant bits from the input to fit the table size. Depending on the table size and the precision, look-up tables can be implemented as BRAM or LUTRAM, which is determined by the HLS compiler.
    For example, on AMD FPGAs, BRAM can be implemented as 18 bit wide by 1024 bit deep true dual-port RAM. Since the variable precision rarely exceeds 18 bits, one BRAM is used for each table of 1,024 elements. In the case of parallel accesses, the HLS compiler may duplicate the BRAM to ensure single-clock-cycle access.
    Empirically, we determine that activation look-up tables consume a small amount of resources and that most activations can be sufficiently approximated with a modest look-up table of $\sim$2048 entries; though heavily quantized models may use even smaller look up tables.

\paragraph{Data flow, scheduling and I/O types:} Models deployed with \hlsfourml target \emph{dataflow} architectures, in which each operation is mapped to a dedicated hardware function synthesized into a separate region of the FPGA fabric. This approach enables pipeline parallelism across the computation graph allowing for low latency and high throughput. In general, dataflow is achieved through directives such as the $\texttt{\#pragma HLS dataflow}$ in \vitishls or tasks in \oneapi. The exact scheduling of layers and data movement between them is, however, determined by the HLS compiler and depends on a number of factors, including the target FPGA platform, clock frequency/uncertainty, and model size. For example, increasing the clock frequency will often lead to higher latency, since the HLS compiler automatically inserts more pipeline stages to better achieve timing closure during place and route.

\begin{figure}
    \centering
    \includegraphics[width=0.85\linewidth]{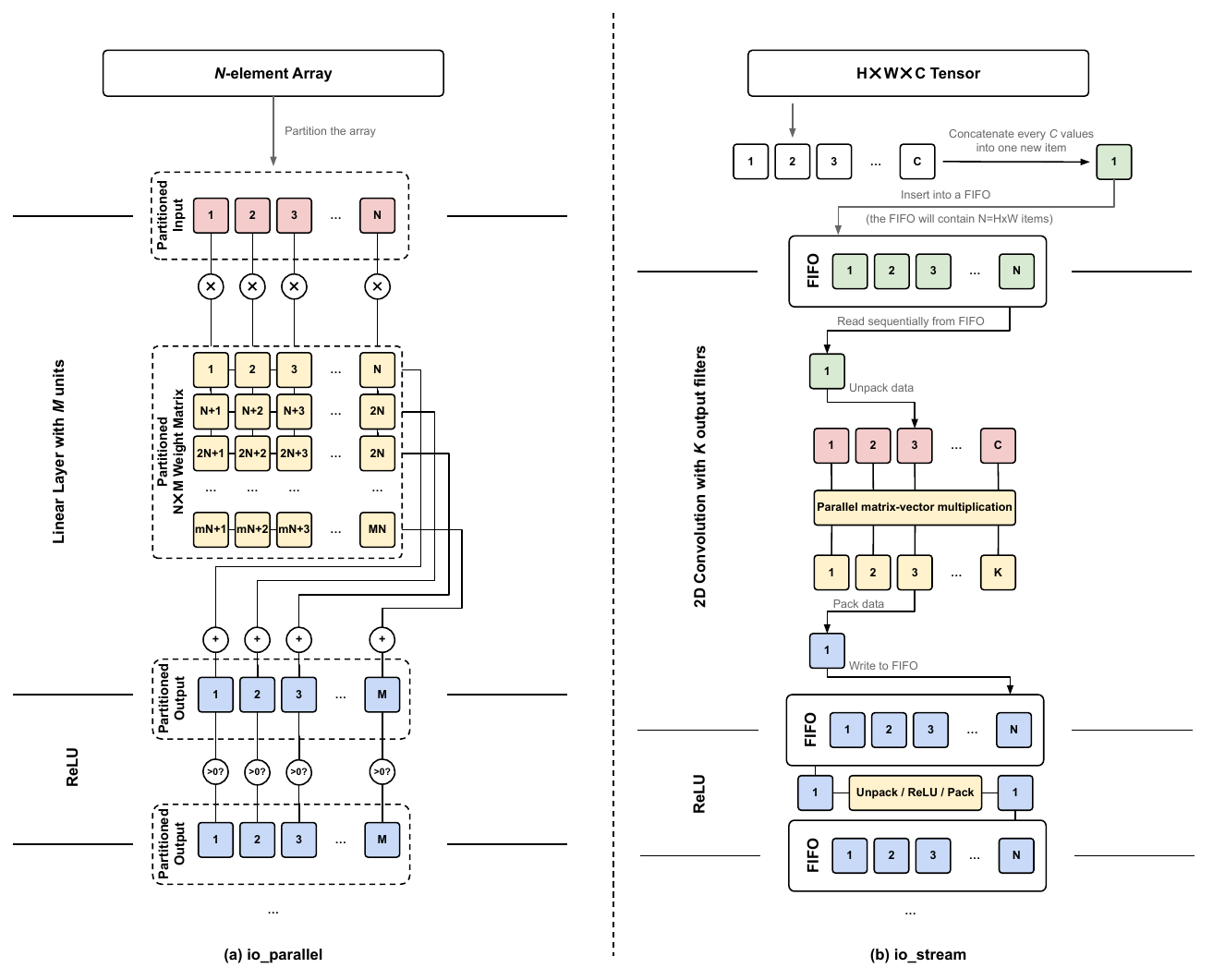}
    \caption{Schematics of the computation of an MLP model implemented using parallel data transfer (left) and a CNN model implemented using streaming data transfer (right). In the Resource strategy, the number of parallel MAC operations executed in each cycle is determined by the RF and PF.
        In the case of the MLP, $\frac{M \cdot N}{\text{RF}}$ multiplications are executed in parallel each clock cycle.}
    \label{fig:io_types}
    \Description{
        Schematic view of the implementation of an MLP using the parallel data transfer on the left side and of a CNN on the right side of the figure.
        For the MLP, vectors and matrices are represented using rectangular boxes containing colored squares representing the values of the individual elements.
        Arrows between the boxes show the data flow, with round symbols indicating the operation.
        We use the times symbol for multiplication, the + symbol for the accumulation, and the greater-equals symbol for a ReLU operation.
        On the left side, the input tensor of size H times W times C is represented as a white rectangle.
        A downward arrow leads to a representation of the individual elements as a series of white squares.
        A rightward arrow represents the concatenation of every C values into a new item, represented by a green square.
        An arrow down from that square than indicates the insertion of that element into a FIFO, which is represented by a rectangular box containing several green squares below the previous symbols.
        A further downward error represents the sequential read from the FIFO.
        A green square at the end of that error shows the extracted element, which is followed by a series of red squares representing the unpacked values.
        These squares are connected by downward lines with a yellow rectangle representing the parallel matrix-vector multiplication in the convolution kernel, from which further lines lead to a series of yellow squares representing the output filters.
        A single downward line leads a blue square representing the packed data, which is then connected to another symbol for a FIFO.
        This is followed by another set of symbols for unpacking, a ReLU, and repacking into a final FIFO.
    }
\end{figure}

The data flow between layers is a crucial part of the design. \hlsfourml supports two types of data transfer between layers: \texttt{io\_parallel} and \texttt{io\_stream} (example in Figure~\ref{fig:io_types}).
The \texttt{io\_parallel} setting directly wires the output of one layer to the input of the next layer, allowing for maximum throughput and minimum latency.
This is the default data transfer type for \hlsfourml, which is suitable for small models with moderate resource usage.
On the other hand, \texttt{io\_stream} uses FIFO buffers between layers.
This is typically used with larger models where some layer requires a significant II, and the data processed by the previous layer need to be buffered before being consumed by the next layer.
Depending on the buffer depth, the HLS compiler may elect to use shift registers or BRAM to implement the FIFOs.
By default, \hlsfourml uses a conservative depth for these buffers, which can result in unnecessary resource overheads.
In the Vivado and Vitis back ends, a special optimizer is implemented to optimize the depth of these FIFOs by performing an RTL co-simulation of the model and recording the maximum occupation of each FIFO buffer~\cite{fifo-depth-optimization}.
The optimizer then sets the final depth for each FIFO to the maximum occupation plus one.

\subsection{IP Core back ends}
Currently, \hlsfourml provides five back ends supporting HLS compilers from AMD/Xilinx, Intel/Altera, and Siemens EDA.
There are two back ends for AMD/Xilinx FPGAs: the Vivado back end, targeting the discontinued \vivadohls, and the Vitis back end, targeting \vitishls 2022.2 or newer.
These two back ends are best supported with the most complete feature set.
They support the most layer types, including fully connected (dense), convolutional, recurrent, many activations, and Einstein summation layers.
Both \texttt{io\_parallel} and \texttt{io\_stream} are supported. All strategies for the CMVM operation are supported, including Latency, Resource, and Distributed Arithmetic.
Due to their popularity, many optimized layers exist for special configurations, such as pointwise, depthwise, and separable convolutions.

Similarly, there are two related back ends for Intel/Altera FPGAs: the Quartus back end, targeting the discontinued \intelhls, and the oneAPI back end, targeting \oneapi 2025.0\footnote{Support for Altera FPGAs was dropped from \oneapi version 2025.1, to be replaced by dedicated Altera software.}. Both \texttt{io\_parallel} and \texttt{io\_stream} are supported in these back ends. In the streaming implementations, the oneAPI back end follows the data flow architecture more closely than the Quartus back end, where individual layers run independently as separate tasks, with handshaking and data transfers between tasks done through the streams implemented as pipes.
Unlike Vivado/Vitis back ends, the Quartus and oneAPI back ends do not provide distinct strategy implementations for the CMVM operations and generally use a resource-optimized approach.

\hlsfourml also provides a back end for \catapulthls, developed in collaboration with Siemens EDA and guided by evolving industry requirements for edge ML applications. \catapulthls is widely used for the development of ASICs, while also supporting FPGA designs, making it a versatile solution for synthesizing C++/SystemC specifications across various hardware targets.
Both Latency and Resource strategies are supported, and the \texttt{io\_type} can be configured as either \texttt{io\_parallel} or \texttt{io\_stream}. For streaming implementations, the Catapult back end supports bottom-up synthesis, leveraging Catapult's Bottom-Up (BUP) flow to enable hierarchical and scalable hardware generation. This methodology enables individual model layers to be synthesized independently without requiring the entire model to be available. Each layer is synthesized into a separate RTL implementation and reusable Catapult library component. This divide-and-conquer approach simplifies design management, supports incremental synthesis (e.g., eliminating the need to re-synthesize the whole model when modifying a single layer), and enhances scalability by reducing memory requirements on the synthesis servers.

\subsection{Accelerator back ends}
The accelerator back ends are extensions of the IP core back ends, which enable end-to-end design-to-deployment flows for specific FPGA/SoC platforms. The accelerator extensions are available for the Vivado and Vitis back ends, with oneAPI currently under development. The Vivado accelerator back end embeds the generated IP core into a larger design containing an DMA core and other necessary utilities for host communication. Beyond IP integration, the back end also facilitates communication between the IP core and a system's memory controller unit (MCU) or the host CPU via a custom device driver. Two different Vitis accelerator back ends are currently being integrated into \hlsfourml. The first uses the \textit{Vitis IP flow}, which closely resembles the Vivado accelerator back end in functionality, primarily targeting the AMD/Xilinx Zynq boards. The second one utilizes the \textit{Vitis System Design Flow}, which enables the creation of complete applications targeting datacenter AMD/Xilinx Alveo accelerators. In this case, the accelerator back end generates a Vitis IP kernel along with a wrapper to facilitate data transfers between the accelerator and the host CPU. This approach allows compiling one or multiple instances of the resulting IP into a binary file that can be directly loaded onto the accelerator card from the host. Additionally, the back end produces multi-threaded host code to manage kernel execution and optimize data movement. This setup enables overlapping PCIe data transfers with kernel computation, effectively multiplexing communication and processing. In practice, this approach has achieved core occupancy levels of up to 96\,\% of the theoretical throughput, as estimated from the II and the kernel's operating frequency. Multi-kernel and multi-accelerator configurations have demonstrated throughput exceeding 1.4\,million inferences per second. \hlsfourml's generic design and interfaces are showcased by its integration with other FPGA shells, such as Coyote v2~\cite{coyote2}, which also allows seamless deployment of applications on Alveo boards.

\section{Community tools for model optimization and co-design}
\label{sec:community_tools}
With the growing need for higher performance and accuracy, machine learning has shifted towards techniques such as quantization and pruning to efficiently deploy models on a variety of hardware.
Such techniques can also be \emph{hardware-aware}, in which the quantization precision or sparsity pattern is determined by the specific computation or memory model of the target device.
Model-hardware co-design often yields the best performance with the lowest drop in accuracy.
Following this trend, the community has proposed a growing number of co-design tools.
We introduce some tools proposed by the community that are supported by \hlsfourml.

\subsection{QKeras}
\qkeras is a widely used QAT library built on top of \kerastwo, developed by Google in collaboration with some of the \hlsfourml authors~\cite{Coelho:2020zfu}.
\qkeras was the first QAT framework to be natively supported in \hlsfourml.
The library supports three quantization types: fixed-point integer, binary, and power-of-two quantization, and provides drop-in replacements for many \kerastwo layers with the corresponding quantized version.
To define a quantized model with \qkeras, one needs to replace the corresponding Keras layer with the corresponding \qkeras counterpart (e.g., \texttt{Conv2D}$\rightarrow$\texttt{QConv2D}, \texttt{Dense}$\rightarrow$\texttt{QDense}), and specify the target quantization type and precision.
AutoQKeras, a hyperparameter tuning tool, may be used to automatically search for the best quantization precision for each layer in the model given a set of constraints on the model size, latency, and accuracy~\cite{Coelho:2020zfu}.
\qkeras has been adopted by various applications, including image classification with quantized CNNs~\cite{fastcnn}, ternary neural networks for network traffic filtering~\cite{ramhorst_dpi}, low-precision models for jet tagging at the Large Hadron Collider (LHC)~\cite{Coelho:2020zfu, ds-fpga}, and muon tracking~\cite{tgc}.

\subsection{High-granularity quantization}
High-granularity quantization (\hgq)~\cite{hgq} is a QAT library built on top of \keras that applies differentiable quantization to weights and activations at a sub-layer granularity.
For models with unrolled CMVM operations, quantization granularity down to the per-parameter level can be used to exploit the FPGA's fine-grained parallelism. \hgq supports only fixed-point integer quantization, treating both binary and power-of-two quantization as special cases of the 1-bit quantization.
Allowing learnable bit-widths to reach 0 bits automatically includes pruning as a special case of quantization. \hgq provides an accurate and differentiable resource usage estimation for the quantized model in the form of effective bit operations (EBOPs), a proxy for computational effort.
This metric is used as a regularization term in the loss function, controlled by a hyperparameter $\beta$, which allows the user to control the trade-off between accuracy and resource usage during training.
On models with unrolled CMVM operations, \hgq demonstrates significant advantages over other model compression techniques, with an on-chip LUT usage reduction of 50\% to 95\%, DSP usage reduction of 50\% to 100\%, and a latency reduction up to 80\% with no inference performance degradation~\cite{hgq, mlpm-fpga}. There are two versions of \hgq: \hgq~1 based on \kerastwo, and \hgq~2 based on \kerasthree.
Both libraries are tightly integrated with \hlsfourml, and no user intervention is required to generate bit-exact HLS designs from the quantized models.

\subsection{Distributed arithmetic}
\label{sec:da4ml}
Distributed arithmetic is a technique for implementing fixed-point multipliers in hardware via shift-and-add operations.
\dafourml~\cite{da4ml} is a library that implements CMVM operations with an adder graph structure. The adder graph in \dafourml is optimized by a set of heuristics to minimize the number of adders weighted by their bit-width, including common sub-expression elimination (CSE).
\dafourml targets unrolled CMVM operations, which constitute the dominant on-chip resource consumption for latency-critical models generated by \hlsfourml.
In contrast to other optimization techniques, \dafourml is an algorithm transformation at implementation time, usually taking only a few seconds, which does not change the model's output by a single bit relative to the corresponding fixed-point implementation.
When applied to models trained with \hgq, \dafourml can achieve further LUT usage reduction of up to $\sim 1/3$ depending on the specific model, while completely eliminating the DSP usage~\cite{da4ml, hgq}.

For some models that require the unrolling of large loops, especially of those contain many multiplication by zero, the HLS code generated directly with \hlsfourml may be unsynthesizable, depending on the HLS compiler. They may, however, be synthesized with \dafourml~\cite{mlpm-fpga, que2025jedi} due to its explicit loop unrolling and resource reduction.
For each CMVM operation implemented with \dafourml, an accurate estimate of LUT usage is also provided.
The \dafourml library is tightly integrated with \hlsfourml, and setting the layer strategy to \texttt{"distributed\_arithmetic"} in the \hlsfourml configuration will enable the distributed arithmetic implementation for the corresponding layers.

\subsection{DSP- and BRAM-aware pruning}
\label{sec:dsp-prune}
Model pruning is an effective way to reduce the resource utilization of a model.
However, the benefit of unstructured pruning is limited for designs where the multipliers are reused, and it is necessary to prune the weights in an organized manner aligned to the specific on-chip primitives used.
The DSP- and BRAM-aware pruning algorithm~\cite{dspprune} is implemented in a submodule of \hlsfourml based on \qkeras with objectives constructed directly corresponding to reductions in on-chip primitive usage. In each case, the algorithm solves a Knapsack problem where weights in the model are assigned an importance value and a hardware cost.
Given a certain resource capacity, the algorithm identifies which weights to prune. As a basic objective, the model can be optimized for sparsity itself in an unstructured approach. However, for a more structured approach, hardware-specific metrics are available.
For FPGAs, these can be registers, DSPs, BRAM, or all of them simultaneously.
For instance, all weights processed by a single DSP can be grouped and pruned together, and hence a DSP block can be removed from the design once a group of weights is pruned.
Depending on the specific application, reductions in DSP usage by factors ranging from 2.2$\times$ to 12.2$\times$ and BRAM usage reductions by factors ranging from 1.4$\times$ to 5.2$\times$ with minimal accuracy degradation have been demonstrated~\cite{dspprune}.

\subsection{Symbolic expression}
\hlsfourml includes a symbolic regression (SR) interface for creating HLS designs based on analytic expressions. SR aims to find a closed-form mathematical expression, containing operators such as $\texttt{exp}(\cdot)$, $\texttt{sin}(\cdot)$, $\texttt{cos}(\cdot)$, that describe the mapping between given input and output data. 
These expressions can be efficiently implemented in hardware, especially if the functions are approximated with look-up tables, making them suitable for low-latency applications. Support for building efficient IPs implementing SR models is included in \hlsfourml on top of the Vivado/Vitis back ends, leveraging the built-in code generation functionality. Specifically, the SR back end supports the translation of expressions represented as strings or SymPy objects, obtained through tools like \textsc{pySR}~\cite{pysr} or SymbolNet~\cite{symbolnet}.
The functions within expressions can be implemented either using the native Xilinx HLS math library or approximated via LUTs, giving users an option to find the desired balance between accuracy, resource usage, and latency. 

\subsection{Surrogate models}
The rapid generation of HLS designs for various models with different hardware configurations opens up the possibility of systematic studies of the design space.
However, waiting for HLS synthesis results is often impractical and lengthy.
An alternative is to use surrogate models, which can accurately predict the model resource utilization and latency.
This can assist in making design choices by providing fast and accurate hardware resource estimates for different model parameter choices.
One example is \textsc{rule4ml}~\cite{rahimifar2024rule4mlopensourcetoolresource}, an open-source tool that uses an MLP surrogate model to predict the resource usage of MLPs generated with \hlsfourml.
Trained on over 15,000 HLS synthesis results covering a wide range of MLP parameter space and \hlsfourml design configurations, it achieves predictions within 10\% of the true synthesis results in $\sim$80\% of test cases.
In a related approach, \textsc{wa-hls4ml}, over 100,000 models were generated, and a graph neural network, lui-GNN, was trained as the surrogate model~\cite{osti_2549315}.

\subsection{Automated neural architecture and hardware co-design}
Neural architecture co-design (NAC)~\cite{Weitz:2025tcj} is a pipeline for neural architecture search and network compression in a two-stage approach based on the ``Once-for-All'' paradigm~\cite{cai2020once} to discover hardware efficient models.
This approach consists of a global search stage that explores a wide range of architectures while considering hardware constraints, followed by a local search stage that fine-tunes and compresses the most promising candidates through QAT and pruning.
Currently, the hardware cost is quantified using bit operations (BOPs), but integration of \hlsfourml resource and latency estimates is underway.
NAC has been demonstrated with Bragg peak finding in materials science and jet classification in HEP, achieving models with improved accuracy, smaller latencies, or reduced resource utilization.

A recent advancement in this direction is MetaML-Pro~\cite{que2025metaml}, a co-optimization framework that automates design space exploration across neural network and hardware abstraction levels. It supports reusable optimization tasks, such as pruning, quantization, and scaling, and enables top-down and bottom-up information flow between software and hardware stages. By integrating tools like \hlsfourml and leveraging Bayesian optimization, MetaML-Pro efficiently balances latency, accuracy, and resource usage, and has been demonstrated on jet tagging models for HEP. Complementary to MetaML-Pro, another recent framework extends co-design principles to incorporate trustworthiness by optimizing not only for accuracy and efficiency but also for uncertainty estimation~\cite{que2025trustworthy}. It introduces Bayesian CNNs with Monte Carlo Dropout and automates trade-offs among predictive accuracy, computational cost, and model calibration, supporting use cases such as safety-critical inference on FPGAs.

\subsection{SoC integration}
The Embedded Scalable Platform (\esp) is an open-source research framework for designing and prototyping heterogeneous system-on-chip (SoC) architectures using a modular, tile-based approach~\cite{mantovani2020agile}. It supports the automated integration of hardware accelerators, processors, memory tiles, and I/O components into an RTL SoC design. ESP integrates with \hlsfourml to enable end-to-end deployment of machine learning models as hardware accelerators~\cite{giri2020esp4ml}. After \hlsfourml generates a synthesizable hardware block from a neural network, ESP wraps it with standard interfaces and connects it to the rest of the SoC, automatically generating testbenches, drivers, and software applications for verification and experimentation. This combined workflow enables the design from a high-level ML model to complete system-level ASIC implementation. FPGA-based emulation and validation are also supported.

\section{User-defined functionality}
\label{sec:extension}

While \hlsfourml supports a wide range of neural networks from a variety of libraries, users may wish to include custom operations in their designs. For layers that are not natively supported by \hlsfourml, users need to provide their own HLS implementation. If these are of wider interest to the community, they can be fully integrated into and contributed to the tool. However, when full integration is not desired, \hlsfourml can be extended via the \texttt{Extension API}, available for both the \kerastwo and \torch front ends, and all supported back ends. The \texttt{Extension API} allows users to leverage most of \hlsfourml's rich infrastructure for other layers, optimizers, and HLS templates, while enabling them to implement the missing operators for their target applications. For this, the front end, IR, and back end must all be extended to support the new layer. On the front end side, this requires a custom layer handler function to parse the properties of the new layer. A class inheriting from the base layer class in the \hlsfourml IR must also be defined. Finally, the back end requires Python templates for the configuration, function calls, as well as the custom HLS code of the new layer.
These components are then registered with \hlsfourml through the \texttt{Extension API}. Optionally, optimizers for the new layer can also be registered. A complete example can be found in the \hlsfourml documentation online\footnote{\url{https://fastmachinelearning.org/hls4ml/advanced/extension.html}}. A  practical example can be found in Reference~\cite{ds-fpga}, where within the context of an interaction network for jet tagging at CERN, the \texttt{Extension API} has been used to implement a custom \kerastwo projection layer, consisting of left- and right-multiplication by graph adjacency matrices. In addition to extending \hlsfourml to support new layers, this mechanism could also be used to provide custom HLS implementations for existing layers, to e.g. optimize the design for a specific application or use different design techniques.

\section{Evaluation}
\label{sec:evaluation}
Given the breadth of possible applications for \hlsfourml, only a fraction of its capabilities can be demonstrated here. The initial performance studies for the different types of networks supported by \hlsfourml can be found in the respective publications~\cite{Duarte:2018ite,fastcnn,linebuffer,Khoda:2022dwz,quartus_rnn_Aad2021,Jiang:2024tkg,Jiang:2024lvg}. Additionally, prior works have shown acceleration of larger CNNs (ResNet- and UNet-like architectures)~\cite{fifo-depth-optimization, d2025edge} as well as large-scale, distributed models over a cluster of FPGAs~\cite{tarafdar2021aigean, lu2024automatic} with \hlsfourml. Here, we instead present quantitative results obtained with the \hlsfourml framework for a variety of benchmark models as new baselines for future works, with the aim of highlighting more recently added capabilities, such as the oneAPI and Catapult HLS back ends. As a community driven project, we also highlight the benefits of the evolving \hlsfourml ecosystem. Therefore, results obtained using the \hgq quantization library and the new DA strategy are presented. These are post-route results of designs generated by \hlsfourml targeting AMD Xilinx UltraScale+ FPGAs (XCVU9P and XCVU13P), or an Altera Agilex7 FPGA. We further highlight the wide range of applications enabled by \hlsfourml by discussing models implemented on ASICs using the Catapult back end, and studies evaluating the performance of the HLS code generated by \hlsfourml.

\subsection{High-level feature jet tagger}
\label{sec:16_feature_jet_tagger}

The high-level jet tagger is a commonly used benchmark for low-latency neural networks on FPGAs~\cite{jet_dataset}.
The task is to classify jets (collimated particle showers at collider physics experiments), into five classes based on their originating particle.
The inputs for each jet are 16 scalar values representing physics-motivated high-level features.
The model architecture employed is an MLP from~\cite{Duarte:2018ite}. The results targeting an AMD XCVU9P FPGA are shown in Table~\ref{tab:jet}.

All models use the \texttt{io\_parallel} data transfer strategy. The models are quantized with \hgq and deployed with both the built-in Latency strategy and the DA strategy provided by \dafourml.
For \hgq trained models, the results shown in different rows are quantized to different bit-widths, regularized by different $\beta$ values.
The results show major improvements when using \hgq over the previous baseline presented in~\cite{qkeras}, shown in the table labeled as \qkeras.
The new DA strategy further reduces the resource usage and latency of the designs.

In addition, we add the maximum achieved clock frequency $\mathrm{F}_\mathrm{max}$ for the models when targeting a 1\,GHz clock period in Table~\ref{tab:jet_fmax} obtained for these designs with the default settings in Vivado and Vitis 2023.2 as a reference.
The results show that the DA strategy may achieve lower latency and comparable or higher $\mathrm{F}_\mathrm{max}$ compared to the Latency strategy. We further show results on a more resource-constrained AMD XC7A35T FPGA in Table~\ref{tab:jet_fmax_xc7a}. In the Latency strategy, the most accurate model failed to place due to insufficient resources, while the DA strategy successfully placed all models, with the most accurate model utilizing more than 90\% of LUTs on the device. This demonstrates that \hlsfourml also enables deployment of models on low-end, resource-constrained FPGAs. Compared to the larger FPGAs, the main difference is reflected in the latency and the clock frequency. Since the AMD XC7A35T is an FPGA of a lower speed grade, the HLS compiler inserts additional register stages to meet timing, leading to higher latency and resource consumption. Additionally, the absolute increase in resources, compared to the larger FPGA, can be attributed to logic duplication during place-and-route, again in order to achieve timing closure.

\begin{table*}[htb]
    \centering
    \caption{Accuracy, resource consumption, latency, and initiation intervals (IIs) of the jet tagging models.
        Resources are reported after out of context place-and-route with an AMD XCVU9P FPGA.
        The clock period used is 5\,ns for all designs shown, except for the design marked with *, which used 7\,ns.
        All designs use no BRAM. The results for QKeras trained models are cited from~\cite{qkeras} and for HGQ cited from~\cite{da4ml}.
    }
    \label{tab:jet}
    \begin{tabular}{ll|cccccc}
        \hline 
        Trainer                       & Strategy & Accuracy & Latency (cc) & DSP        & LUT            & FF            & II \\
        \hline

        {QKeras}~\cite{dspprune}$^*$  & Latency  & 76.3\%   & 15 [105 ns]  & 5,504      & 175            & 3,036         & 2  \\

        {QKeras}~\cite{que2025metaml} & Latency  & 76.1\%   & 10 [50 ns]   & 13,042     & 70             & N/A           & 1  \\

        \hline
        HGQ                           & Latency  & 76.9\%   & 5 (23.6 ns)  & 55 [0.8\%] & 13,303 [1.1\%] & 2,374 [0.1\%] & 1  \\
        HGQ                           & Latency  & 76.5\%   & 5 (23.1 ns)  & 30 [0.4\%] & 6,715 [0.6\%]  & 1,348 [0.1\%] & 1  \\
        HGQ                           & Latency  & 75.9\%   & 3 (12.7 ns)  & 15 [0.2\%] & 3,044 [0.3\%]  & 641 [0.0\%]   & 1  \\
        \hline
        HGQ                           & DA       & 76.9\%   & 5 (23.4 ns)  & 0          & 11,978 [1.0\%] & 2,117 [0.1\%] & 1  \\
        HGQ                           & DA       & 76.5\%   & 4 (18.0 ns)  & 0          & 6,067 [0.5\%]  & 1,178 [0.0\%] & 1  \\
        HGQ                           & DA       & 75.9\%   & 3 (12.1 ns)  & 0          & 2,891 [0.2\%]  & 651 [0.0\%]   & 1  \\

        \hline
    \end{tabular}
\end{table*} 

\begin{table*}[htb]
    \centering
    \caption{Accuracy, resource consumption, latency, and $\mathrm{F}_\mathrm{max}$ of the jet tagging models trained in HGQ~\cite{hgq} when targeting 1\,GHz clock. Resource reported are after out of context place-and-route with an AMD XCVU9P FPGA (part number: \texttt{xcvu9p-flga2104-2L-e}). All designs use no BRAM and have an II of one cycle.}
    \label{tab:jet_fmax}
    \begin{tabular}{ll|cccccc}
        \hline 
        Trainer & Strategy & $\mathrm{F}_\mathrm{max}$ & Accuracy & Latency (cc) & DSP        & LUT            & FF             \\
        \hline
        HGQ     & Latency  & 739 MHz                   & 76.9\%   & 37 (50.1 ns) & 57 [0.8\%] & 14,624 [1.2\%] & 21,365 [0.9\%] \\
        HGQ     & Latency  & 705 MHz                   & 76.5\%   & 31 (44.0 ns) & 30 [0.4\%] & 7,656 [0.6\%]  & 11,513 [0.5\%] \\
        HGQ     & Latency  & 694 MHz                   & 75.9\%   & 30 (43.2 ns) & 17 [0.2\%] & 3,765 [0.3\%]  & 5,845 [0.2\%]  \\
        \hline
        HGQ     & DA       & 713 MHz                   & 76.9\%   & 28 (39.3 ns) & 0          & 12,465 [1.1\%] & 18,004 [0.8\%] \\
        HGQ     & DA       & 719 MHz                   & 76.5\%   & 23 (32.0 ns) & 0          & 6,297 [0.5\%]  & 9,524 [0.4\%]  \\
        HGQ     & DA       & 756 MHz                   & 75.9\%   & 21 (27.8 ns) & 0          & 2,886 [0.2\%]  & 4,667 [0.2\%]  \\
        \hline
    \end{tabular}
\end{table*} 

\begin{table*}[htb]
    \centering
    \caption{Accuracy, resource consumption, latency, and $\mathrm{F}_\mathrm{max}$ of the jet tagging models trained in HGQ~\cite{hgq}. Resource reported are after out of context place-and-route with an AMD XC7A35T FPGA (part number: \texttt{xc7a35tlcsg325-2L}). All designs use no BRAM and have an II of one cycle.}
    \label{tab:jet_fmax_xc7a}
    \begin{tabular}{ll|cccccc}
        \hline 
        Trainer & Strategy & $\mathrm{F}_\mathrm{max}$ & Accuracy & Latency (cc)                     & DSP         & LUT             & FF              \\
        \hline
        HGQ     & Latency  & -                         & 76.9\%   & \multicolumn{4}{c}{Place failed}                                                   \\
        HGQ     & Latency  & 187 MHz                   & 76.5\%   & 53 (284.0 ns)                    & 30 [33.3\%] & 11,091 [53.3\%] & 24,649 [59.3\%] \\
        HGQ     & Latency  & 182 MHz                   & 76.0\%   & 45 (246.7 ns)                    & 17 [18.9\%] & 5,565 [26.8\%]  & 11,330 [27.2\%] \\
        \hline
        HGQ     & DA       & 172 MHz                   & 76.9\%   & 58 (337.4 ns)                    & 0           & 19,467 [93.6\%] & 36,995 [88.9\%] \\
        HGQ     & DA       & 185 MHz                   & 76.5\%   & 50 (270.1 ns)                    & 0           & 10,692 [51.4\%] & 20,407 [49.1\%] \\
        HGQ     & DA       & 176 MHz                   & 75.9\%   & 46 (261.1 ns)                    & 0           & 5,455 [26.2\%]  & 9,952 [23.9\%]  \\
        \hline
    \end{tabular}
\end{table*} 

On the same task, we also processed a \qkeras trained model with 76.9\% accuracy using the oneAPI back end, and evaluated its performance using \oneapi 2025.0 and \quartus 23.1. The results are given in Table~\ref{tab:jet_altera}. Given the model's small size, we continue using \texttt{io\_parallel} for data transfer between the layers, and we fully unroll the layers. For this compiler, we use different compiler flags for different optimization targets. We found that the lowest latency can be obtained when requesting the lowest target frequency simultaneously with the latency optimization target. As this model has an II of one under this configuration, its throughput is proportional to the clock frequency given a continuous data stream. Therefore, it is possible to trade off between latency and throughput by adjusting the target frequency.

\begin{table*}[htb]
    \centering
    \caption{Resource consumption, latency, and $\mathrm{F}_\mathrm{max}$ of the jet tagging \qkeras models using \oneapi. All designs have an accuracy of 76.9\%. Resources reported are after out of context place-and-route, as reported by \quartus, with an Altera Agilex7 FPGA (part number: \texttt{AGFB014R24A2E2V}). The latency is measured first input to first output, reported both in clock cycles and in nanoseconds if running at $\mathrm{F}_\mathrm{max}$. The optimization target is an option given to the \texttt{icpx} compiler, unrelated to the \hlsfourml strategy. The default option, obtained by not explicitly specifying an optimization target, optimizes for maximum frequency, and turns on hyper-optimized handshaking. All designs shown have negligible MLAB, RAM, and DSP usage (less than 0.01\%) and an II of one.}
    \label{tab:jet_altera}
    \begin{tabular}{ll|cccccc}
        \hline
        Opt. Target & Target $\mathrm{F}_\mathrm{max}$ & $\mathrm{F}_\mathrm{max}$ & Latency (cc) & ALM            & ALUT           & FF             & \\
        \hline
        Latency     & 200 MHz                          & 341 MHz                   & 21 [62 ns]   & 11,434 [2.3\%] & 21,471 [2.2\%] & 8,864 [0.45\%] & \\
        Latency     & 480 MHz                          & 614 MHz                   & 50 [81 ns]   & 12,085 [2.5\%] & 22,551 [2.3\%] & 31,243 [1.6\%] & \\
        Default     & 480 MHz                          & 782 MHz                   & 90 [115 ns] & 13,576 [2.8\%] & 24,319 [2.5\%] & 40,051 [2.1\%] & \\
        \hline
    \end{tabular}
\end{table*} 

\subsection{Street View House Numbers classifier}
\label{sec:svhn}

Classification of the house numbers in images from the Street View House Numbers (SVHN) dataset~\cite{svhn} is a common benchmark for computer vision tasks.
We show the results from~\cite{hgq} and~\cite{fastcnn}, where both used the same model architecture as in~\cite{fastcnn}: a CNN with 3 convolutional layers, 3 max pooling layers, and 2 fully connected layers.
The detailed description of the architecture can be found in both works.
For \hgq trained models, the results shown in different rows are quantized to different bit-widths, regularized by different $\beta$ values.
The results are shown in Table~\ref{tab:svhn}.
All designs shown are using $\text{PF}=1$ (each convolution kernel is applied once per clock cycle) and \texttt{io\_stream}.
Compared to the previous results obtained in~\cite{fastcnn} with \hlsfourml, the new results use significantly fewer on-chip resources while retaining the same accuracy.
As in the previous section, we find that the DA strategy offers improved performance compared to the Latency one, most notably by eliminating all DSP usage, while reductions in LUTs, FFs, and latency are less pronounced. However, this advantage might not transfer to cases where high variable precision is required, and all cases with an RF larger than one, which are not supported in the DA strategy.

\begin{table*}[htb]
    \centering
    \caption{Accuracy, resource usage, and latency of the SVHN classifier models. Reported resource usage after out of context place-and-route with an AMD XCVU9P FPGA. The clock period used is 5 ns. The results for \qkeras trained models are cited from~\cite{fastcnn}, and the \hgq trained models are cited from~\cite{hgq}.}
    \label{tab:svhn}
    \begin{tabular}{ll|ccccccc}
        \hline
        Trainer & Strategy & Accuracy & Latency (cc)       & DSP          & LUT (k)     & FF (k)      & BRAM         & II (cc) \\
        \hline
        \qkeras & Latency  & 94.\%    & 1,035 [5.2 $\mu$s] & 174 [2.54\%] & 111 [9.4\%] & 33 [1.4\%]  & 67.0 [3.1\%] & 1,030   \\
        \qkeras & Latency  & 88.\%    & 1,059 [5.3 $\mu$s] & 72 [1.05\%]  & 48 [4.1\%]  & 15 [0.63\%] & 32.5 [1.5\%] & 1,029   \\
        \hline
        HGQ     & Latency  & 93.9\%   & 1,050 [5.3 $\mu$s] & 58 [0.85\%]  & 69 [5.8\%]  & 28 [1.2\%]  & 32.0 [1.5\%] & 1,029   \\
        HGQ     & Latency  & 93.1\%   & 1,061 [5.3 $\mu$s] & 30 [0.44\%]  & 47 [4.0\%]  & 21 [0.89\%] & 28.0 [1.3\%] & 1,029   \\
        HGQ     & Latency  & 91.9\%   & 1,058 [5.3 $\mu$s] & 15 [0.22\%]  & 40 [3.4\%]  & 18 [0.76\%] & 23.5 [1.1\%] & 1,029   \\
        \hline
        HGQ     & DA       & 93.9\%   & 1,045 [5.2 $\mu$s] & 0            & 53 [4.5\%]  & 20 [0.85\%] & 32.0 [1.5\%] & 1,029   \\
        HGQ     & DA       & 93.1\%   & 1,045 [5.2 $\mu$s] & 0            & 37 [3.1\%]  & 15 [0.63\%] & 28.0 [1.3\%] & 1,029   \\
        HGQ     & DA       & 91.9\%   & 1,045 [5.2 $\mu$s] & 0            & 31 [2.6\%]  & 14 [0.59\%] & 23.5 [1.1\%] & 1,029   \\
        \hline
    \end{tabular}
\end{table*}

\subsection{Particle-based jet tagger}
\label{sec:150_particle_jet_tagger}

The particle-based jet tagger is a more challenging benchmark for low-latency neural networks on FPGAs~\cite{jet_dataset}.
Similar to the high-level jet tagger, the task is to classify jets into five classes based on their originating particle. However, instead of using the 16 high-level features, which are unlikely to be available in real-time, the raw particle information is fed into the model.
Up to 150 particles are available for each jet, ordered by their transverse momentum.
Each particle has 16 features, and up to 64 particles are used in the model (zero-padded if less than 64 particles are available).
The model architecture is an MLP-Mixer tailored for this task, and the detailed description can be found in~\cite{mlpm-fpga}.
The results are reproduced following the same setup as in~\cite{mlpm-fpga} with updated template functions and the latest \hlsfourml version.
The results are shown in Table~\ref{tab:particle_jet_tagger}.
The models shown on different rows are quantized to different bit-widths, regularized by different $\beta$ values. The \texttt{io\_parallel} configuration is used for all models, and only the DA strategy is used, 
as the Latency strategy fails to converge in timing and to meet the initiation interval constraints.
This is likely due to the overly large, but highly sparse kernel matrices used in the MLP-Mixer architecture, where the HLS compiler's heuristic failed to implement the design properly.

\begin{table*}[htb]
    \centering
    \caption{
        Accuracy, resource usage, and latency of the particle-based classifier models in HGQ. Reported resource usage after out of context place-and-route with an AMD XCVU13P FPGA.
        The clock period used is 5\,ns.
        All HGQ trained models are using the MLP-Mixer architecture from~\cite{mlpm-fpga}. Models without DA failed to meet timing and II constraints.
        All designs use no DSPs or BRAM and have an II of one cycle.
    }
    \label{tab:particle_jet_tagger}

    \begin{tabular}{ll|cccc} 
        \hline
        Trainer & Strategy & Accuracy & Latency (cc) & LUT (k)     & FF (k)      \\
        \hline
        HGQ     & DA       & 81.4\%   & 13 [65 ns]   & 126 [7.3\%] & 26 [0.76\%] \\
        HGQ     & DA       & 81.0\%   & 13 [65 ns]   & 68 [3.9\%]  & 13 [0.38\%] \\
        HGQ     & DA       & 80.3\%   & 12 [60 ns]   & 51 [3.0\%]  & 11 [0.33\%] \\
        \hline
    \end{tabular}
\end{table*}

\subsection{MNIST classification}
\label{sec:mnist}

Classification of images of handwritten digits from the Modified National Institute of Standards and Technology (MNIST) database~\cite{mnist} is a common benchmark for computer vision tasks.
The task is to classify handwritten digits from 0 to 9.
The model architecture employed a single layer MLP with 128 hidden units and 10 output units. 
We trained three models with different quantization levels with HGQ, all using the \texttt{io\_parallel} approach for data transfer between layers, and the results are shown in Table~\ref{tab:mnist_classifier}. Only the DA strategy can be applied to these models, as
when using the Latency strategy, the sparse $768\times128$ kernel matrix fails to unroll properly, leading to HLS synthesis failures.

\begin{table*}[htb]
    \centering
    \caption{Accuracy, resource usage, latency, and initiation intervals of the MNIST classifier models in HGQ~\cite{hgq}. Reported resource usage after out of context place-and-route with an AMD XCVU13P FPGA. The clock period used is 5 ns. All HGQ trained models are using a single layer MLP. Models without DA failed to synthesize. All designs use no DSPs or BRAM and have an II of one cycle.}
    \label{tab:mnist_classifier}

    \begin{tabular}{ll|cccc}
        \hline
        Trainer & Strategy & Accuracy & Latency (cc) & LUT           & FF             \\
        \hline
        HGQ     & DA       & 97.7\%   & 2 [10 ns]    & 9671 [0.56\%] & 2,039 [0.06\%] \\
        HGQ     & DA       & 97.3\%   & 2 [10 ns]    & 6890 [0.40\%] & 1,672 [0.05\%] \\
        HGQ     & DA       & 97.0\%   & 2 [10 ns]    & 5040 [0.29\%] & 1,349 [0.04\%] \\
        \hline
    \end{tabular}
\end{table*} 

\subsection{Smart pixels}
\label{sec:smartpixels}

In~\cite{parpillon2024smartpixelsinpixelai,yoo2024smart}, we demonstrated the implementation of a neural network-based filter for low-momentum charged particle tracks in ``smart pixel'' sensors, targeting on-chip data reduction to cope with the even higher data rates expected in future HEP experiments. Using the Catapult back end, we translated a compact quantized classifier (\textasciitilde 1,200 trainable parameters) from \qkeras into synthesizable C++ and generated fully combinatorial RTL code for integration in 28\,nm CMOS. The design was implemented in conjunction with the analog front end, using less than $1\,\mu\text{W}$ / pixel for digital logic within a total power budget below $1\,\text{W}/\text{cm}^2$.
The HLS-driven flow enabled faster design cycles, efficient exploration of precision and parallelism, and ultimately achieved a bandwidth reduction between 54.4\% and 75.4\%. 

Instead of transmitting full pixel cluster information, we further aim to compress the data into a small number of variables that describe the kinematics of the traversing particle.
Using a mixture density network implemented directly in the pixel readout electronics, we can predict the track angle and hit position along with the associated uncertainties.
This additional layer of intelligent data reduction enables even lower bandwidth and downstream computational complexity while preserving critical physics information.
The most recent model is a five-layer quantized network, which in \hlsfourml is implemented as a pipeline with an initiation interval of one, a latency of two clock cycles, using the \texttt{io\_parallel} and Latency strategies.
The model consists of two convolutional layers followed by three dense layers (\textasciitilde 3,000 trainable parameters).
The first convolutional layer is implemented as a depthwise-separable convolution to minimize the number of parameters and reduce computation, which was added to the Catapult back end in \hlsfourml for this project.
Using 2D convolutions, this model has an estimated area of 0.81\,mm$^2$ in 28\,nm CMOS, while a 1D convolutional variant achieves comparable functionality with a significantly reduced area of 0.39\,mm$^2$.
Ongoing design space exploration is focused on further minimizing area while preserving predictive performance and hardware feasibility.

\subsection{Quality of \hlsfourml results}
\label{sec:rtl_vs_hls}
While the ease of use of HLS frameworks like \hlsfourml is a significant benefit, the quality of results is a critical factor for adoption.  
Comparing the resource usage, latency, and throughput of \hlsfourml generated designs against hand-mapped Verilog designs for the same models, the HLS-synthesized designs can benefit from more sophisticated constant folding in the HLS compiler compared with the FPGA vendor's standard synthesis tools, the exploitation of carefully constructed Verilog codes that better invoke the back end tools, and other optimizations~\cite{CarolineICCAD,CarolineThesis}. 
The \hlsfourml flow can achieve superior performance compared to hand-mapped designs in some cases, while being up to $2\times$ worse in the worst case. Similar results are demonstrated in~\cite{YilinThesis} for automated synthesis flows targeting embedded AI engines (AIEs).
When comparing \hlsfourml designs automatically mapped to the FPGA fabric to software designs mapped automatically to the VLIW processors embedded in some AMD Versal devices, the \hlsfourml designs were generally superior to the AIE implementations in resource usage, latency, and throughput, while being up to $2\times$ worse in the worst case.

\section{Applications of \hlsfourml}
\label{sec:motivation}

\hlsfourml was originally developed for HEP applications at CERN and has since been widely adopted for a variety of use cases. The ATLAS and CMS detectors~\cite{The_ATLAS_Collaboration_2008, The_CMS_Collaboration_2008} at the CERN LHC~\cite{Lyndon_Evans_2008} deploy a data filtering system on FPGAs that decides in real time which collisions to store persistently, reducing data volumes from approximately 100\,TB/s to about 10\,GB/s with a typical latency constraint of $\sim$10\,$\mu$s.
As of now, three algorithms have been deployed using \hlsfourml. The first is an MLP to assign muon transverse momentum and displacement in triggers for unconventional signatures~\cite{CMS:2023gfb}. The AXOL1TL~\cite{CMS-DP-2024-059,CMS-DP-2023-079} and CICADA~\cite{CMS-DP-2024-121,CMS-DP-2023-086} algorithms are variational and convolutional autoencoders, respectively, trained to detect anomalies~\cite{Govorkova:2021utb} in the detector data with different levels of preprocessing. Similar applications with \hlsfourml are being considered for other HEP experiments and related applications~\cite{Hao:2025mwz,Giasemis:2025jhj,MehdiRahimifar:2024hkb}.

The need for real-time data analysis makes the deployment of ML models on FPGAs desirable in other scientific environments as well.
For example, \hlsfourml was used to demonstrate that CNN models deployed on FPGAs could meet latency requirements that would enable self-triggering of radio antennas for cosmic ray detection~\cite{Dorosti:2025ugq}.
Similarly, the necessity of real-time control of sensitive systems makes these techniques suitable for quantum computing applications, where \hlsfourml has been used in the development of MLP models for the control~\cite{Bhat:2024lnc} and read out~\cite{DiGuglielmo:2025zod} of qubits.
As an example of an application in accelerator control~\cite{StJohn:2020bpk}, a demonstrator system was designed using \hlsfourml employing an Intel Arria 10 FPGA-based board to monitor and correctly attribute beam losses in the Fermilab Recycler Ring~\cite{shi2023mlbasedrealtimecontroledge}.
Similarly, for active feedback control in magnetic confinement fusion devices an inference latency of 7.7\,$\mu$s and end-to-end latency of 17.6\,$\mu$s was achieved, enabling the tracking of magnetohydrodynamic modes in the plasma at over 100 kfps~\cite{10.1063/5.0190354}.
Potential applications also extend to radiation safety, where a compact, light-weight, low-power radiation detector has been developed in which an MLP model is deployed on an Artix-7 Series FPGA using \hlsfourml~\cite{GammaNeutron}.

\hlsfourml has also expanded out from scientific computing into diverse domains including automotive and space applications. It has been used to prototype a quantized CNN for semantic segmentation in self-driving cars~\cite{linebuffer}, achieving a latency of $\sim$5ms while using less than 30\% of the available resources on a Xilinx ZCU102 board.
Power efficiency constraints are a key concern for deploying ML models on satellites and other spacecraft. \hlsfourml is being used in the development of models for environmental monitoring of the earth using satellites~\cite{edgespaice}.
In another signal processing application, \hlsfourml has enabled the real-time analysis of microwave sensing data for food contamination using an MLP~\cite{Stitic:2024hra}, achieving a latency of $27\,\mu$s on an AMD/Xilinx Kria K26 FPGA.
\hlsfourml was used to deploy a combined CNN and RNN model (C-RNN) for accurate classification of artificial surface textures with a latency of $6.21\,\mu$s~\cite{SurfaceTextureRNN}. Demonstrating its usefulness in biomedical applications,
\hlsfourml was also used to prototype a 1D-CNN for arrhythmia classification from ECG waves on a system-on-chip (SoC), with a power consumption of 1.655\,W~\cite{10399904}. Another example is the classification of lymphocyte cells, where \hlsfourml was used to demonstrate a 12x reduction in inference latency compared to GPUs when deploying the classifier on a Euresys frame grabber FPGA~\cite{islam2025realtimecellsortingscalable}. \hlsfourml was used to accelerate a ternary model for detecting malicious executables in network packets with a latency of 44\,ns, while maintaining the 100\,Gbps network throughput~\cite{ramhorst_dpi}.
In a speech recognition application, a 1D CNN model was deployed using \hlsfourml on a Pynq-Z2 FPGA, achieving a performance improvement of 6$\times$ to 12$\times$ compared to GPUs and CPUs~\cite{SpokenKeywordSpotting}. Given the growing community, use cases, and contributions, continued development and support of \hlsfourml as an open-source, high-performance and modular platform for neural network acceleration on FPGAs is imperative.

\section{Challenges and limitations}
\label{sec:limitations}
\hlsfourml's place in the ecosystem as an interface between a growing set of training libraries for deep learning models and HLS tools presents a variety of challenges for its continued development. The large number of front and back ends already available or in active development will further increase the maintenance burden and requires increased community engagement. The heterogeneous support of features in the different front and back ends outlined in Section~\ref{sec:overview} is therefore likely to persist, depending on the level of community interest in the different tools. This is exacerbated by the fact that novel implementations derived using \hlsfourml in the context of research projects are not always easy to generalize and contribute back to the tool. This limits the usability of \hlsfourml for some tasks, especially if the user is restricted to use one of the lesser developed back ends.

Problems can also arise from HLS tools, especially for larger models where the complexity might overwhelm the HLS compiler, leading to long, or in extreme cases, failed HLS synthesis. 
Given the integration of \hlsfourml with a large number of external tools and products, maintaining full functionality of the tool and high performance requires extensive validation effort. Different HLS compiler versions frequently differ (e.g. supported pragmas may change), and changes in performance for the same HLS code between versions are common. This requires regular adjustments and reoptimization of the HLS templates in \hlsfourml. In extreme cases, \hlsfourml has to react to the discontinuation of whole tools, such as support for Altera FPGAs being dropped from the Intel \oneapi compiler starting with version 2025.1.

\section{Related work}
\label{sec:related_work}
The acceleration of ML/AI inference on FPGAs is a rapidly evolving field, with numerous proprietary and open-source solutions proposed. Broadly, FPGA neural network accelerators can be classified as (1) software-controllable, overlay-based or (2) custom hardware generators~\cite{boutros2024field, 9444092}. Overlay architectures implement a fixed microarchitecture in FPGA fabric, which is controllable from software similar to CPUs and GPUs. Conversely, custom hardware accelerators generate optimized model-specific designs, typically by creating dedicated hardware implementations for individual layers or operators in the model. In this section, we provide a brief overview of the existing tools and their features, focusing mainly on platforms with custom hardware generation, since \hlsfourml falls into this category. More in-depth overviews of neural network acceleration on FPGAs, including overlay architectures, can be found in surveys~\cite{boutros2024field, 9444092}.

\begin{table*}[hbt]
    \newcommand{\tworows}[2]{\begin{tabular}[c]{@{}c@{}}#1 \\ #2\end{tabular}}
    \newcommand{\mfend}{\tworows{Multiple}{ML frameworks}}
    \newcommand{\mbend}{\tworows{Multiple}{hardware vendors}}
    \newcommand{\quant}{\tworows{Arbitrary}{Quantization}}
    \newcommand{\yes}{\faCheckSquare}
    \newcommand{\no}{\faSquareO}
    \newcommand{\partially}{\faPlusSquareO}
    \caption{Comparison of different tools for the translation of ML models to FPGA designs.}
    \label{tab:related_work}
    \resizebox*{\textwidth}{!}{
        \begin{tabular}{ccccccccc}
            Framework                                     & \mfend & \mbend & \quant     & CNN        & RNN  & Transformers & Open-source & Maintained \\
            \hline
            Vitis AI~\cite{amd2024vitisai}                & \yes   & \no    & \no        & \yes       & \yes & \yes         & \yes        & \yes       \\
            Intel FPGA AI Suite~\cite{intel2024fpgasuite} & \yes   & \no    & \no        & \yes       & \yes & \yes         & \no         & \yes       \\
            EdgeCortex MERA~\cite{edgecortix2025mera}     & \yes   & \no    & \no        & \yes       & \yes & \yes         & \yes        & \yes       \\
            \hline
            FP-DNN~\cite{7966671}                         & \no    & \yes   & \yes       & \yes       & \yes & \no          & \no         & N/A        \\
            DeepBurning~\cite{7544251}                    & \no    & \yes   & \no        & \yes       & \yes & \no          & \yes        & \no        \\
            DeepHLS~\cite{9294881}                        & \no    & \yes   & \yes       & \yes       & \no  & \no          & \yes        & \no        \\
            PipeCNN~\cite{8280160}                        & \no    & \yes   & \yes       & \yes       & \no  & \no          & \yes        & \no        \\

            DnnWeaver~\cite{dnnweaver:micro16}            & \no    & \yes   & \no        & \yes       & \no  & \no          & \yes        & \no        \\
            NNGen~\cite{takamaeda2017nngen}               & \yes   & \yes   & \yes       & \yes       & \no  & \no          & \yes        & \no        \\

            fpgaConvNet~\cite{7544745}                    & \yes   & \no    & \yes       & \yes       & \no  & \no          & \yes        & \partially \\
            HPIPE/H2PIPE~\cite{9974441,H2PIPE}            & \no    & \yes   & \yes       & \yes       & \no  & \no          & \no         & N/A        \\
            MASE~\cite{cheng2023fast,zhangmase}           & \no    & \yes   & \yes       & \yes       & \no  & \yes         & \yes        & \yes       \\
            cgra4ml~\cite{cgra4ml}                        & \no    & \yes   & \yes       & \yes       & \no  & \yes         & \yes        & \yes       \\

            chisel4ml~\cite{10456782}                     & \yes   & \yes   & \yes       & \yes       & \no  & \no          & \yes        & \yes       \\
            tinyHLS~\cite{tinyHLS}                        & \no    & \yes   & \partially & \partially & \no  & \no          & \yes        & \yes       \\
            FINN~\cite{blott2018finn,finn}                & \yes   & \no    & \yes       & \yes       & \no  & \yes         & \yes        & \yes       \\
            \hline
            \hlsfourml~\cite{Duarte:2018ite}              & \yes   & \yes   & \yes       & \yes       & \yes & \partially   & \yes        & \yes
        \end{tabular}
    }
    \medskip

    \yes: Supported; \partially: Partially supported; \no: Not supported. \\ First group: commercial products; second group: research projects. \\ ``Multiple ML frameworks'' corresponds to a framework being able to ingest models from multiple different frameworks; however, if a framework supports \onnx/\qonnx, it automatically supports multiple frameworks, since \kerastwo and \torch can be exported to \onnx. ``Arbitrary quantization'' refers to the ability to accelerate arbitrarily quantized models; rather than supporting only fixed quantization strategies (e.g., \texttt{INT8}, \texttt{FP16}).
\end{table*}

Table~\ref{tab:related_work} summarizes the projects described in this section and their capabilities. Commercial platforms include AMD \textsc{Vitis AI}~\cite{amd2024vitisai} and Intel \textsc{FPGA AI Suite}~\cite{intel2024fpgasuite}, both of which can convert neural network models from multiple deep learning frameworks. The MERA~\cite{edgecortix2025mera} compiler from EdgeCortex converts models from \torch, \tensorflow, and \onnx into dataflow graphs that can be deployed on their dynamic neural architecture IP cores. However, neither framework extends to multiple hardware vendors nor supports arbitrary quantization.

On the research side, early platforms include \textsc{FP-DNN}~\cite{7966671} and \textsc{DeepBurning}~\cite{7544251}, which use RTL to generate hardware designs, making them portable across platforms. HLS-based platforms include \textsc{DeepHLS}~\cite{9294881} and \textsc{PipeCNN}~\cite{8280160} that support MLP and CNN models. Neither of these four platforms support multiple front ends, and to our knowledge, none are actively maintained. \textsc{fpgaConvNet}~\cite{venieris2016fpgaconvnet,venieris2017latency,venieris2018mapping} focuses specifically on translating large-scale CNNs defined in \torch or \onnx into HLS code for \vivadohls. Similarly, \textsc{HPIPE} provides a workflow to create optimized CNN designs starting from \tensorflow models for various FPGAs using RTL design. Both frameworks include support for arbitrary, per-layer quantization. To support even larger models, \textsc{H2PIPE}~\cite{H2PIPE} extends \textsc{HPIPE} with designs that utilize off-chip memory. For even larger models, such as transformers, two notable platforms are \textsc{MASE}~\cite{cheng2023fast,zhangmase} and \textsc{cgra4ml}~\cite{cgra4ml}. \textsc{MASE}~\cite{cheng2023fast,zhangmase} supports quantization and conversion of \torch models into FPGA designs using templates in SystemVerilog, making it portable across hardware platforms. Notably, it directly enables QAT, with a strong focus on transformer models and LLMs. Similarly, \textsc{cgra4ml}~\cite{cgra4ml} focuses on accelerating large neural networks by reusing processing units across layers, including support for MLPs, CNNs, and transformers. Unlike \hlsfourml, \textsc{cgra4ml} does not generate dataflow designs and stores weights in off-chip memory. Neither \textsc{MASE} nor \textsc{cgra4ml} support multiple front ends: \textsc{MASE} is compatible with \torch and \textsc{cgra4ml} with \qkeras. \textsc{chisel4ml}~\cite{10456782} enables the translation of MLPs and CNNs from \qonnx into fully parallelized, dataflow designs supporting arbitrary quantization via the Chisel hardware description language. In general, performance comparable to that of \hlsfourml was reported~\cite{10456782}. On the other hand, \textsc{tinyHLS}~\cite{tinyHLS} generates \texttt{Verilog} code enabling MLPs and one-dimensional CNNs defined in \kerastwo to be deployed on FPGAs. \textsc{FINN}~\cite{blott2018finn,finn} is an open-source project by AMD/Xilinx that allows translating models defined in \brevitas or \qonnx into hardware designs targeting AMD/Xilinx FPGAs. Like \hlsfourml, it targets a dataflow architecture with fine-grained quantization control for low-latency, high-throughput application, and supports a wide range of neural network architectures.

\section{Conclusions and outlook}
\label{sec:conclusion}
FPGAs are uniquely suited for efficient and flexible deployment of neural networks in low-latency, low-power environments. However, traditional implementations of these models in RTL are complex and require specialized knowledge. A large number of tools and platforms aim to aid users in the translation of their models into FPGA designs, removing the need to re-implement common building blocks. Among these, \hlsfourml is an open-source platform that stands out for its modular design and extensible nature. \hlsfourml supports all major deep learning libraries and different HLS compilers, enabling deployment across a wide range of FPGA vendors as well as ASIC design via integration with \catapulthls. Furthermore, \hlsfourml supports a wide and growing range of model architectures and provides a convenient interface to include custom layers in the design. The FPGA designs available in \hlsfourml are easily customizable for different objectives, such as latency or resource usage. Hardware-software co-design is crucial for achieving accurate and high-performance designs, with pruning and QAT as the most common techniques. \hlsfourml supports models trained with most common QAT libraries and has integrated support for hardware-aware compression techniques. Thus, \hlsfourml plays a central role in a growing ecosystem for optimized deployment of neural networks on FPGAs. Additionally, it can act as a platform that supports research in novel model-hardware co-design techniques.

\hlsfourml is developed by an active, diverse community of researchers, continuously expanding its capabilities and features. At its core, the \hlsfourml community strives to keep the platform modular and extensible, continuously adding new front and back ends.
For example, a Flax/JAX front end and a \texttt{SmartHLS} compiler back end targeting Microchip PolarFire FPGAs are in development. In parallel, to support NanoXplore radiation-tolerant FPGAs, the Catapult HLS library will be extended, or alternatively, integration with the PandA Bambu HLS toolchain \cite{ferrandi2021bambu} will be added to the back ends.
Significant emphasis is being placed on supporting and developing novel co-design libraries to maximize performance while minimizing resource and latency requirements.
On the hardware design side, support for transformer architectures are active area of development, with a first implementation being available via the \hgq~2 framework.
A variation of the transformer architecture optimized for computational efficiency on point cloud data using locality sensitive hashing (HEPT)~\cite{Miao:2024oqy} is also being developed in \hlsfourml.
Another promising direction for future development is the integration of AMD AI engines, and \hlsfourml may take advantage of this architecture alongside its existing support for FPGA designs. Following \hlsfourml's recent integration with Coyote~\cite{coyote, coyote2}, an open-source shell with rich networking services and collective communication~\cite{accl}, future work may extend \hlsfourml to support distributed inference, by leveraging these services. Such distributed inference would be particularly suitable for larger models that cannot fit on a single FPGA, e.g., transformers.

\begin{acks}
  We acknowledge the importance of the Fast Machine Learning community for the development of this project. We especially thank Paolo D'Alberto, Duen-Jeng (DJ) Wang and Luciano Lavagno from AMD, as well as Yaman Umuroglu, Michaela Blott, and the overall FINN team, for helpful insights and facilitating technical discussions.
  This work was partly supported by the Accelerated AI Algorithms for Data-Driven Discovery (A3D3) Institute under U.S. National Science Foundation (NSF) Grant No. PHY-2117997. This work was supported by FermiForward Discovery Group, LLC under Contract No. 89243024CSC000002 with the U.S. Department of Energy (DOE), Office of Science, Office of High Energy Physics. Work done by Imperial College is funded by the Science and Technology Facilities Council (STFC) grant ST/W000636/1 and EPSRC (grant numbers UKRI256, EP/V028251/1, EP/N031768/1, EP/S030069/1, and EP/X036006/1). Enrico Lupi, Dimitrios Danopoulos, and Vladimir Lon\v{c}ar are funded by the Eric \& Wendy Schmidt Fund for Strategic Innovation through the CERN Next Generation Triggers project under grant agreement number SIF-2023-004. Nicolò Ghielmetti is supported by the European Research Council (ERC) under the European Union’s Horizon 2020 research and innovation program (Grant Agreement No. 101135358). This work was partially supported by DOE ASRSP GCFA Grant ID: SP0062070 and NSF POSE Phase II Award 2303700. Dylan Rankin is supported by the DOE, Office of Science, Office of High Energy Physics Early Career Research program under Award No. DE-SC0025324. Javier Duarte is additionally supported by the Research Corporation for Science Advancement (RCSA) under grant \#CS-CSA-2023-109, Alfred P. Sloan Foundation under grant \#FG-2023-20452, and DOE, Office of Science, Office of High Energy Physics Early Career Research program under Award No. DE-SC0021187. Santosh Parajuli and Mark Neubauer are additionally supported by the U.S. Department of Energy, Office of Science, Office of High Energy Physics, under contract number DE-SC0023365. Chang Sun is partially supported by the NSF ACCESS Grant number PHY240298, DOE, Office of Science, Office of High Energy Physics grant under Award No. DE-SC0011925. Jennifer Ngadiuba and Chang Sun are partially supported by the DOE Office of Science, Office of High Energy Physics ``Designing efficient edge AI with physics phenomena'' Project ({DE-FOA-0002705}). Thea K. Aarrestad is supported by the Swiss National Science Foundation Grant No.~PZ00P2\_201594. Ryan Forelli gratefully acknowledges support from the Ryan Fellowship and the International Institute for Nanotechnology at Northwestern University. Benjamin Ramhorst gratefully acknowledges AMD for the donation of the Heterogeneous Accelerated Compute Cluster (HACC) at ETH Zurich, which was partially used for the development and testing of the work presented in this paper.
\end{acks}

\bibliographystyle{ACM-Reference-Format}
\bibliography{hls4ml}

\end{document}